\documentclass[a4paper,reqno]{amsart}
\usepackage[english]{babel}
\usepackage[T1]{fontenc}
\usepackage{amsmath,amssymb,amsfonts,amsthm}
\usepackage{hyperref,comment}
\usepackage{slashed}
\usepackage{graphicx,color}
\usepackage[dvipsnames]{xcolor} 
\usepackage[bbgreekl]{mathbbol}
\usepackage{mathtools}
\usepackage[all,cmtip]{xy}

\usepackage{soul}
\usepackage[colorinlistoftodos,textsize=footnotesize]{todonotes}

\mathtoolsset{showonlyrefs}

\newtheorem{definition}{Definition}

\newtheorem{remark}[definition]{Remark}

\newcommand{\intl}[1]{\int\limits_{#1}}

\newcommand{\B}{\mathcal{B}}

\newcommand{\Hsf}{\mathsf{H}}
\newcommand{\BRST}{\textsf{\tiny BRST}}

\newcommand{\ph}{\varphi}
\newcommand{\RR}{\mathbb{R}}
\newcommand{\Hcal}{\mathcal{H}}
\newcommand{\Fcal}{\mathcal{F}}
\newcommand{\Ecal}{\mathcal{E}}

\newcommand{\Wcal}{\mathcal{W}}
\newcommand{\oph}{\overline{\ph}}

\newcommand{\Ci}{\mathcal{C}^\infty}

\newcommand{\scri}{\mathcal{I}}

\newcommand{\bt}{{\boldsymbol{t}}}
\newcommand{\bl}{{\boldsymbol{l}}}

\newcommand{\loc}{\mathrm{loc}}
\newcommand{\rad}{\mathrm{Rad}}
\newcommand{\her}{\mathrm{Her}}

\newcommand{\be}{\begin{equation}}
\newcommand{\ee}{\end{equation}}

\title{Asymptotic symmetries in the BV-BFV formalism}
\author{Kasia Rejzner}
\address[K. Rejzner]{Department of Mathematics, University of York, Heslington, York YO10 5DD, United Kingdom}
\email{kasia.rejzner@york.ac.uk}

\author{Michele Schiavina}
\address[M. Schiavina]{Institute for Theoretical Physics, ETH Z\"urich,  Wolfgang-Pauli-Str. 27, 8093 Z\"urich, Switzerland and Department of Mathematics, ETH Z\"urich, R\"amistrasse 101, 8092 Z\"urich, Switzerand}
\email{micschia@phys.ethz.ch}

\begin{document}

\maketitle

\begin{abstract}
We show how to derive asymptotic charges for field theories on manifolds with ``asymptotic'' boundary, using the BV-BFV formalism. We also prove that the conservation of said charges follows naturally from the vanishing of the BFV boundary action, and show how this construction generalises Noether's procedure. Using the BV-BFV viewpoint, we resolve the controversy present in the literature, regarding the status of large gauge transformation as symmetries of the asymptotic structure. We show that even though the symplectic structure at the asymptotic boundary is not preserved under these transformations, the failure is governed by the corner data, in agreement with the BV-BFV philosophy.
We analyse in detail the case of electrodynamics and the { interacting scalar field, for which we present a new type of duality to a sourced two-form model}. 
\end{abstract}

\tableofcontents

\section*{Introduction}
Asymptotic symmetries for field theories in the presence of ``boundaries at infinity'' have received great attention recently, after they have been shown to be related to Weinberg soft theorems \cite{Weinberg65}. The asymptotic structure of quantum electrodynamics (QED) and general relativity (GR) has  also  been studied in a number of earlier works (see for example \cite{Ash81,Ash81b,AS81,Ash81c,Ashtekar87} for GR and quantum gravity and \cite{Her95,r05903,r05892,r05907,Sta99,Sta02,r05873,r19355,r05884,Her16} for QED). 
A great scientific effort has been devoted to this topic in the last decade, showing how asymptotic charges are expected to arise in a host of scenarios, including the crucial examples of general relativity \cite{CL14,HLMS15}, electrodynamics \cite{HMPS14,CL15,Strominger} and even scalar field theories \cite{CCM,CC,CFHS}. More abstractly, the question of whether a gauge symmetry can \emph{become global}, and hence present observable charges, is a relevant one for both theoretical modelling and experimental probing of fundamental theories. 

One could also ask whether the conservation laws for the asymptotic charges in question really arise from symmetries of the theory, i.e. transformations of fields that preserve the action functional as well as the canonical symplectic form. This concern has been raised in \cite{Her16}, where the asymptotic charge of QED and its conservation is derived as a consequence of field equations, rather than the Noether procedure applied to \textit{large gauge transformations} (LGT), in contrast to \cite{Strominger}. Here, by large gauge transformations we mean: transformations of the fields whose parameters have nonvanishing asymptotics.  Such transformations are shown to relate QED theories in different gauges and they do not preserve the canonical symplectic structure on (asymptotic) Cauchy data \cite{Her16}. In \cite{DybWe19} it was shown that quantum theories in different gauges could be unitarily inequivalent, which would mean that a transformation that changes the gauge  does not actually preserve the quantum theory. This leads to the conundrum: \textit{in what sense are the large gauge transformations symmetries of the theory?}

In this paper, we solve this conundrum by employing a framework called BV-BFV --- due to Cattaneo, Mnev and Reshetikhin \cite{CMR1}.
For a manifold with boundary, the BV-BFV framework is a combination of the Batalin--Vilkovisky (BV) approach to the quantisation of a Lagrangian field theory associated to the ``bulk'' of said manifold, and the Batalin--Fradkin--Vilkovisky (BFV) approach to its corresponding Hamiltonian formulation, naturally associated to the boundary\footnote{BFV provides a resolution of the reduced phase space of the theory, i.e. of the locus defined by canonical constraints modulo symmetries.} \cite{BV77,BV81,BF83}.

We adapt the BV-BFV framework to the case of ``boundary at infinity'', to which we associate the asymptotic scaling limit of a theory assigned to the boundary of a scaled finite region. From such extended data we extract information on asymptotic symmetries and charges. At the classical level, while we agree with the observation of \cite{Her16} that large gauge transformations have to relate theories in different gauges and do not preserve the canonical (boundary) symplectic structure, we are able to show how they can be interpreted as \textit{extended symmetries}. Indeed, failure of gauge invariance of the relevant boundary structures is to be expected, and is interpreted as structural corner data.

The first advantage of the BV-BFV setting, when discussing the interpretation of LGTs, is the model-independence and flexibility of the framework, which allows for a direct generalisation of Noether's analysis of charges. As a consequence, we are able to reproduce the formulas from the literature on both sides of the controversy and point out where the interpretational discrepancies stem from. This is not surprising, since the BV-BFV data carry information about both the  Lagrangian symmetries and the behaviour of solutions to the equations of motion. Thus, after identifying the asymptotic charge with the \emph{BV-BFV boundary action} at infinity (see below), we can interpret it both from the point of view of Noether charges (the interpretation favoured e.g. by \cite{Strominger,CL15,CE17}) and from the view-point of field equations (relating to the interpretation of \cite{Her16}). We show that, assuming appropriate fall-off conditions for the fields, one can easily read off the correct expressions for asymptotic charges from the BV-BFV data naturally associated to a theory on a manifold with boundary, and prove their conservation. 

Our results agree with the literature in examples of electrodynamics and the massless scalar field. In particular, we compare the results on electromagnetic asymptotic charges presented in \cite{Her95,r05903,r05892}  with the investigations of \cite{Strominger,CL15,CE17}. Using the same procedure, we derive the soft charges for the scalar field, compare them with those derived in \cite{Her95,CCM,CC}, and show their conservation.

To recover the hard charges for scalar fields we propose a new kind of duality between a sourced scalar field and a sourced two-form model (Section \ref{Sec:hardscalar}). To our knowledge, this duality was not considered before, and recovers the usual duality in the sourceless limit. We are then able to completely recover asymptotic charges for the sourced scalar field from the BFV boundary action associated to the sourced dual two-form model, evaluated on asymptotic data. 

While our result is similar in spirit to the analysis of \cite{CFHS}, we disagree on the definition and the need of what they call ``large gauge transformations'': shifts by closed-but-not-exact forms (this differs from the nomenclature we adopt, see above). Instead, we derive scalar asymptotic charges and their conservation by implementing the (reducible) symmetry of the dual two-form model in the BV-BFV formalism. Then, in Section \ref{s:LGTnoBV} we show that transformations of the type employed by \cite{CFHS} do not yield a well-defined BV structure, making their interpretation and relevance harder to pin down.

The second advantage of the BV-BFV approach, in this context, is the possibility to encode gauge invariance anomalies of relevant data in terms of cohomological data one codimension higher, effectively setting up a bulk-to-boundary or boundary-to-corner correspondence. This allows for a straightforward generalisation of the notion of \textit{symmetry} of a field theory, where boundary and corner terms are not to be discarded, but serve rather as higher codimension structural data. Our point of view relates to descent equations \cite{ZUMINO1985477,MSZ} (see the recent perspective \cite{MSW}), but is extended to a full symplectic and cohomological description of higher codimension data, for which a quantisation scheme exists \cite{CMR2}.

While the interplay of Lagrangian symmetries and equations of motion  is central to the BV philosophy, bulk-boundary correspondences are at the core of the BV-BFV framework. By combining these two philosophies we propose a systematic approach to the calculation of asymptotic charges, and a new interpretation thereof as extended symmetries. We discuss our new interpretation of conserved asymptotic charges of QED in Sections~\ref{EMcorners} and \ref{sec:dis}, where we compare to the literature and argue how this resolves the interpretational conundrum.

In a broader context, our long-term goal is building a bridge between a systematic approach to the quantisation of gauge theories in the presence of boundary,\footnote{In this case the ``boundary'' is at infinity.} such as the BV-BFV formalism, and asymptotic quantisation. The latter is an idea dating back to \cite{Ash81} to address infrared problems  in QED and in quantum gravity (related to the masslessness of the photon and graviton, and the long-range nature of the interaction)  by analysing the structure of asymptotic observables at null infinity.

The long-range character of the electromagnetic interaction manifests itself in the classical theory via Gauss' law. In  quantum theory, implementability of Gauss' law, together with the assumption that observables should be local, leads to the conclusion that the electric flux at space-like infinity is superselected (i.e., different configurations of the flux label different unitarily-inequivalent representations of the net of local algebras \cite{r05889}). Alternatively, one can implement Gauss' law in the quantum algebra, where the fluxes are not superselected, paying the price of giving up the locality \cite{r05892}. Other phenomena related to the long-range character of electrodynamics include breaking of the Lorentz group and the infraparticle problem \cite{MS86}. The latter means that the electron's spectrum is not point-like, since the electron has to be considered together with the cloud of low energy (infrared) photons accompanying it. This fact, in different guises, can be understood as a necessity for ``dressing'' charged particles, as discussed, for example, in \cite{DF16,Dyb17} and references therein. 

This paper is the first step towards developing a unified framework for quantisation of theories with boundaries and theories with long-range degrees of freedom, in the spirit of perturbative algebraic quantum field theory \cite{FR,FR3}. 
The framework we develop in this paper for the construction of classical asymptotic charges is general enough to treat a broad spectrum of theories. One only needs to specify the dynamics, the boundary/appropriate ``infinity'', and the behaviour of fields at this boundary/infinity. Then our extended BFV machinery returns the correct conserved charge. Although we use mainly the language of \cite{CMR1}, the translation to the classical framework of \cite{FR} is straightforward.

A third main advantage of the BV-BFV approach is a direct access to a flexible quantisation scheme. The axioms that a classical field theory with boundary needs to satisfy are the starting point of the procedure presented in \cite{CMR2}, which has been tested on a variety of field theories (e.g. $BF$ theory \cite{CMR2,CMRBF}, Yang--Mills theory in dimension 2 \cite{IM}, and split Chern--Simons theory \cite{Cattaneo2017}).

The asymptotic adaptation of BV-BFV quantisation, and the precise relation to \cite{FR3} is still work in progress. However, we expect to phrase Weinberg's soft theorems in this language, relating the quantum master equation, modified by the presence of boundaries, to Ward identities.

The application of our procedure to other scenarios like nonabelian Yang--Mills, Chern--Simons and $BF$ theories is expected to be straightforward. The case of General Relativity (GR) in the Einstein--Hilbert (EH) formalism, whose BV-BFV structure for finite boundaries was investigated in \cite{Schia,CSEH}, will be studied in a further publication. In space-time dimension 3 the BV-BFV construction of GR in vielbein variables --- often called Palatini--Cartan formalism --- was presented in \cite{CaSc}, while for its 4-dimensional analogue a crucial obstruction was found in \cite{Schia,CSPCH}. On the other hand, a BFV structure has been recently worked out from the reduced phase space of Palatini--Cartan theory in dimension $n\geq3$ \cite{CSPCHcl,CCS}, independently from the BV theory in the bulk, and a BV-BFV-compatible formulation of tetradic gravity has been given in \cite{CCS2020b}. We do expect asymptotic symmetries in this formulation to be easier to compute than their Einstein--Hilbert counterpart.

In Section \ref{s:preliminaries} we review the basics of the BV-BFV approach to field theories on manifolds with boundary, showing how it reproduces and extends Noether's analysis of conserved charges (Section \ref{s:noethercomparison}),  and state the necessary geometric conventions for the remainder of the paper. We discuss the extension of the BV-BFV formalism to corners in Section \ref{Corners}, and introduce the notion of extended symmetries in Section \ref{extgauge}. Finally, we introduce two descriptions of classical asymptotic data: the one in Section~\ref{sec:Rsl} is based on the approach of Herdegen (see e.g. \cite{Her16}) and the other one, introduced in Section~\ref{Sect:coordinates}, is used by \cite{CFHS,CCM,CC}.

Section \ref{s:electrodynamics} concerns the asymptotic symmetries of electrodynamics (ED): firstly without matter fields and next in the presence of  scalar matter. We show how one obtains the asymptotic charges from the BFV data associated to ED, seen as abelian Yang--Mills theory, once appropriate fall-off conditions on fields are imposed. This agrees with \cite{Her16,CE17,CL15}. In Section \ref{EMcorners} we complement the analysis of asymptotic charges with a discussion of the symplectic structure of ED, and its behaviour under large gauge transformations. We show how the role played by corner terms (and their BV-BFV interpretation) is key for the resolution of the interpretational conundrum around LGT's. 

In Section \ref{Sect:Scal} we apply the same procedure to the (free) two-form model, dual to a (free) scalar field on-shell, and recover the soft asymptotic charges for scalar field theory, through the BV-BFV analysis of its associated dual model. We extend this construction to the sourced scenario, and propose a modified duality between a scalar and a two-form model in Section \ref{Sec:hardscalar}. Finally, we argue how constant shifts and --- dually --- symmetries generated by closed but not exact forms do not yield a BV structure in Section \ref{s:LGTnoBV}.

\section{Preliminaries}\label{s:preliminaries}

\subsection{Fields and functionals}\label{sec:fieldsandfunctinals}
We start this section by defining some geometrical structures, which we need in order to formulate the BV-BFV formalism. For more information on infinite dimensional differential geometry see for example \cite{Michor,Neeb06}.

Let $M$ be a compact manifold with boundary (later on we will generalise this to non-compact manifolds by imposing appropriate falloff conditions on fields). 
In the simplest case of a field theory, classical field configurations are modelled as sections of some (potentially graded) vector bundle. In general (e.g. in the case of gravity) the space of field configurations is instead an infinite dimensional manifold. In this paper we only consider the simplest situation, but all the structures introduced here generalise straightforwardly.

Let $E\xrightarrow{\pi} M$ be a, possibly graded, vector bundle over $M$, and denote its space of smooth sections by $\mathcal{E}\doteq \Gamma(M,E)$, equipped with the standard Fr\'echet topology. 
We can define classical observables as functionals on $\Ecal$, i.e. smooth maps in $\Ci(\Ecal,\RR)$. Smoothness is understood in the sense of Bastiani calculus \cite{Bas64} (see also \cite{BDGR} for a review). Most importantly, functional (variational) derivatives of functionals in this framework are distributional sections. More precisely: for $F\in \Ci(\Ecal,\RR)$, we have $F^{(n)}(\ph)\in\Gamma'(M^n,E^{\boxtimes n})$, where $\boxtimes$ is the exterior tensor product of vector bundles and the prime denotes the strong dual (topological dual equipped with the strong topology).

Among all functionals, important role is played by \textit{local functionals}. These are those which can be written as
\[
F(\ph)=\int \omega(j_x^k(\ph))\,,
\]
where $\omega$ is a top form on $M$ that depends only on the finite jet $j_x^k(\ph)$ of the field configuration $\ph$ at point $x$ (intuitively, $j_x^k(\ph)$ is the value of $\ph$ and its derivatives up to order $k$ at  point $x$, see \cite{And} for more on jet spaces in field theory). Let $\Ci_{\loc}(\Ecal,\RR)$ denote the space of local functionals.

We can consider the tangent space $T_{\ph}\Ecal$ of $\Ecal$ at a given point $\ph\in\Ecal$ and we notice that $T_{\ph}\Ecal\cong \Ecal$. Let $T\Ecal$ denote the tangent bundle of $\Ecal$. Vector fields are understood as smooth sections of this bundle and we observe $\Gamma(T\Ecal)\cong\Ci(\Ecal,\Ecal)$. 

We define the cotangent bundle using the strong dual, meaning that $T^*_\ph\Ecal\doteq\Ecal'$, which is the space of distributional sections of $E$. Let $E^*$ be the dual bundle of $E$ and $\Ecal^*$ its space of smooth sections. We use the notation $\Omega^1(\Ecal)\equiv \Gamma(T^*\Ecal)$ for 1-forms on $\Ecal$, i.e. smooth maps from $\Ecal$ to $\Ecal'$.

Denote $\Ecal^!\doteq \Ecal^*\otimes\mathrm{Dens}$, where $\mathrm{Dens}$ is the space of densities on $M$. We note that there is a natural pairing between $\Ecal$ and $\Ecal^!$, which we denote by $\left<.,.\right>$. Using this pairing we identify elements of $\Ecal^!$ with distributional sections in $\Ecal'$, so there is a natural inclusion $\Ecal^!\subset \Ecal'$. 

Analogously to local functionals we can define also local forms $\Omega_\loc^1(\Ecal)$ and local vector fields $\Gamma_\loc(T\Ecal)$. All this also generalises to multivector fields $\bigwedge^\bullet \Gamma(T\Ecal)$ and $n$-forms $\Omega^\bullet(\Ecal)$.

For convenience of computation and in order to make contact with the physics literature, we introduce a formal notation for functionals, forms and vector fields on $\Ecal$. First of all, we note that an important role is played by evaluation functionals on $\Ecal$. Let $x\in M$ and fix a basis $e_\alpha$ on the fibre of $E$, so that $\ph(x)=\ph^\alpha e_\alpha$. We define $\Phi^\alpha_x\in\Ci(\Ecal,\RR)$ by
\be\label{eq:evaluation}
\Phi^\alpha_x(\ph)\doteq \ph^\alpha(x)\,.
\ee
We can write local functionals in terms of those evaluation functionals. By the common abuse of notation, we will often use the notation $\ph^\alpha(x)$ instead of $\Phi_x^\alpha$, i.e. we use the same symbol for points in $\Ecal$ and coordinate functions on $\Ecal$. From now on we will also suppress the index $\alpha$ in all the summations.

Vector fields, as derivations on $\Ci(\Ecal,\RR)$, can be formally written as:
\[
X=\int X_\ph(x)\frac{\delta}{\delta \ph(x)}\,,
\]
where $X\in\Gamma(T\Ecal)$, in the sense that $X$ acting on $F$ (i.e. the Lie derivative of $F$ with respect to $X$) is
\[
\mathcal{L}_X F(\ph)=\int X_\ph(x)\frac{\delta F}{\delta \ph(x)}\,.
\]
Here $X_\ph\in\Ecal$, so $X$ is a map from $\Ecal$ to $\Ecal$, as required. The objects $\frac{\delta}{\delta \ph^\alpha(x)}$ are vector-field-valued distributions and we can think of them as forming a ``basis'' for the vector fields (in the same sense as $\Phi_x^\alpha$ form a ``basis'' for local functionals). In physics literature these are called \textit{antifields} and we will denote them  by $\Phi^\ddagger_x$ or $\ph^\ddagger(x)$. Later on, we will consider the odd cotangent bundle $T^*[-1]\Ecal$ and we want to treat $\Phi_x$ and $\Phi^\ddagger_x$ on equal footing, as formal generators. We will then also consider derivatives with respect to these formal generators and denote them by $\frac{\delta}{\delta\ph(x)}$ and $\frac{\delta}{\delta\ph^\ddagger(x)}$ respectively.

As for 1-forms, we denote by $\delta F$ the 1-form obtained from a functional $F\in \Ci(\Ecal,\RR)$ by taking the derivative, i.e. $\delta F(\ph)\in\Ecal'$ and for $\psi\in\Ecal$,
\[
\left<\delta F(\ph),\psi\right>\doteq  \lim_{t\rightarrow 0}\frac{1}{t} (F(\ph+t\psi)-F(\ph))\,,
\]
where the pairing $\left<.,.\right>$ is the natural dual pairing between $\Ecal$ and $\Ecal'$. 

The map $\delta$ from $\Ci(\Ecal,\RR)$ to $\Omega^1(\Ecal)$ is identified as the de Rham differential. It is extended to $n$-forms by the graded Leibniz rule. The Lie derivative $\mathcal{L}$ is also extended to $n$-forms on $\Ecal$ by the formula
\[
\mathcal{L}_X=\iota_X \delta - \delta\iota_X\,.
\]

In particular, for the evaluation functional $\Phi_x^\alpha$, we can define $\delta\Phi_{x}^\alpha$. The corresponding 1-form-valued distribution will be denoted by $\delta\Phi^\alpha$ and by some abuse of notation also $\delta\ph^\alpha$. 

The insertion of a vector field $X$ into 1-form $\delta\Phi_x^\alpha$ results in the following functional:
\[
(\iota_X \delta\Phi_x^\alpha)(\ph)=X_\ph^\alpha(x)\,,
\]
so we can think of 1-form-valued distributions $\delta\Phi^\alpha$ as dual to vector-field-valued distributions $\Phi_x^\ddagger\equiv \frac{\delta}{\delta\ph(x)}$, with the dual pairing given by:
\[
\left<\delta\varphi^\alpha(x), \frac{\delta}{\delta \ph^\beta(y)} \right>=\delta^\alpha_{\beta}\,\delta(x-y)\,.
\]
We can then write arbitrary 1-forms as
\[
\Omega(\ph)=\int {\Omega_\ph}(x) \delta\Phi_x\,,
\]
so the insertion of a vector field into a 1-form can be expressed in terms of the above dual pairing as:
\begin{multline*}
(\iota_X\Omega)(\ph)=\left<\int \Omega_\ph(x)\delta\Phi_x,\int X_\ph(y)\Phi_y^\ddagger\right>\\=\int\int \Omega_\ph(x) X_\ph(y)\delta(x-y)=\int \Omega_\ph(x) X_\ph(x)\,.
\end{multline*}
These considerations naturally generalize to multivector fields and $n$-forms. A degree-$k$ multivector field $X$ can be expressed in terms of antifields as
\[
X=\int X(x_1,\dots,x_k)\Phi_{x_1}^\ddagger\wedge \dots\wedge  \Phi_{x_k}^\ddagger\,,
\]
where the product $\wedge$ is (graded) antisymmetric. The \emph{antifield number} is the polynomial degree of these multivector fields. Similarly, a degree-$n$ form can be written as:
\[
\Omega=\int Q(x_1,\dots,x_k)\delta\Phi_{x_1}\wedge \dots\wedge  \delta\Phi_{x_k}\,,
\]

Note that polyvector fields can be interpreted as functions on the graded manifold $T^*[-1]\Ecal$ (the \textit{odd cotangent bundle}). If $\Ecal$ is just degree 0, then $T^*[-1]\Ecal=\Ecal[0]\oplus \Ecal'[-1]$, so the functions on $T^*[-1]\Ecal$ are identified with $\Ci(\Ecal,\RR)\hat{\otimes}\bigwedge^\bullet \Ecal\cong \bigwedge^\bullet\Gamma(T\Ecal)$, as required. For the precise definition of the completed tensor product $\hat{\otimes}$ and all the topologies involved, see e.g. \cite{Book}. As mentioned before, the antifields can be understood as odd generators, and elements of $\bigwedge^\bullet\Gamma(T\Ecal)$ are functions of both fields $\ph$ and the antifields $\ph^\ddagger$. In this sense, one can also define left and right derivatives with respect to $\ph^\ddagger$ and repeat the discussion presented in this section to introduce vector fields and $n$-forms on $T^*[-1]\Ecal$. Among those, the special role is played by the odd Poisson bivector
\[
\Pi=\int \frac{\delta}{\delta \ph(x)^\ddagger}\wedge \frac{\delta}{\delta \ph(x)}\,,
\]
which defines the \textit{antibracket}
\[
\{F,G\}=\iota_{\Pi}(\delta F\wedge \delta G)\,,
\]
for $F,G\in \bigwedge^\bullet\Gamma(T\Ecal)$ such that this is well defined (e.g. for $F,G$ local). By dualization, one can write this bracket also in terms of the odd symplectic form
\be\label{eq:BVform}
\Omega =\int \delta \Phi_x^\ddagger\wedge \delta \Phi_x
\ee
by means of
\[
\{F,G\}=\iota_{X_F}\iota_{X_G}\Omega\,,
\]
where $X_F$ is the vector field obtained by contraction of $\Pi$ with the 1-form $\delta F$, similarly with $X_G$. Note that $\delta$ in \eqref{eq:BVform} is the de Rham differential on $T^*[-1]\Ecal$, rather than on $\Ecal$. Finally, we define the canonical one-form $\alpha$ to be
\begin{equation}\label{BVform}
\Omega=\delta \int  \Phi_x^\ddagger\wedge\delta\Phi_x = \delta\alpha.
\end{equation}

\subsection{BV-BFV formalism}\label{Sect:BVBFV}
A classical field theory on a manifold with boundary phrased in the BV-BFV formalism \cite{CMR1,CMR2} is described by two sets of data, one assigned to the bulk manifold and one to the boundary, together with an appropriate map between the two. To the bulk manifold $M$ one associates BV data
$$
    (\mathcal{F} , \Omega , S , Q )
$$
composed of
\begin{enumerate}
    \item A $(-1)$-symplectic graded manifold $(\mathcal{F} , \Omega )$. 
    \item A degree $0$ action functional
    \item An odd vector field $Q $ on $\mathcal{F}$  of degree $1$ with the cohomological property $[Q ,Q ]=0$.
\end{enumerate}
In simplest cases (e.g. Yang-Mills theory), $\mathcal{F}$ is the odd cotangent bundle $T^*[-1]\mathcal{E}$ of some graded manifold $\mathcal{E}$ (containing the classical fields in degree zero and the ghosts in degree one). Functionals on $T^*[-1]\mathcal{E}$ are identified with polyvector fields on $\Ecal$, as defined in Section \ref{sec:fieldsandfunctinals}. { More generally, one can always identify (noncanonically) any $(-1)$-shifted symplectic manifold with an odd cotangent bundle \cite{SchwarzBV}, so we will use it as a universal model.}

To a boundary $\partial M$ one assigns (exact) BFV data
$$
    (\mathcal{F}^\partial , \Omega^\partial , S^\partial , Q^\partial )
$$
similarly composed of
\begin{enumerate}
    \item An exact $(0)$-symplectic graded manifold $(\mathcal{F}^\partial , \Omega^\partial =\delta\alpha^\partial )$, where $\delta$ denotes the de Rham differential on the space of local forms, 
    \item A degree $1$ local action functional $S^\partial $ on $\mathcal{F}^\partial $,
    \item An odd vector field $Q^\partial $ on $\mathcal{F}^\partial $ of degree $1$ with the property: $[Q^\partial ,Q^\partial ]=0$.
\end{enumerate}

The BV-BFV construction connects the BV data associated with the bulk to the BFV data associated with the boundary by means of a map 
\begin{equation}\label{BVBFVmap}
    \pi \colon\mathcal{F} \longrightarrow \mathcal{F}^\partial,
\end{equation}
and the following relations hold
\begin{subequations}\label{BVBFVeqts}\begin{align}\label{BVBFV}
    \iota_{Q }\Omega  
        &= \delta S  
            + \pi ^* \alpha^\partial  \\\label{mCME}
    \frac12 \iota_{Q }\iota_{Q }\Omega  
        &= \pi ^* S^\partial \\\label{boundaryBVBFV}
    \iota_{Q^\partial }\Omega^\partial  
        &= \delta S^\partial \\\label{boundaryCME}
    \frac12 \iota_{Q^\partial }
        \iota_{Q^\partial }\Omega^\partial  
        &= 0
\end{align}\end{subequations}

\begin{remark}
Observe that if $M$ has no boundary one defines BV data such that Equations \eqref{BVBFV} and \eqref{mCME} hold without the corrections coming from the boundary. In particular, in that case, $Q $ is the Hamiltonian vector field of $S $ and equation \eqref{mCME} becomes the Classical Master Equation. 
\end{remark}

If the BV theory is constructed from the data of a classical field theory with (gauge) symmetries, the degree-zero part of $\mathcal{F} $ and $S $ coincide with the classical data $(F_{\text{cl}} ,S_{\text{cl}})$, the space of classical fields and the classical action functional. The BV-complex, given by $\mathfrak{BV}^\bullet\coloneqq (C^\infty(\mathcal{F} ), Q )$ is a combination of the Koszul--Tate resolution of the critical locus of $S_{\text{cl}}$ and of the Chevalley--Eilenberg complex for Lie algebra actions. In this language, the space of on-shell invariant functionals is given by the zeroth cohomology group $H^0(\mathfrak{BV})$.

The BFV data represents the reduced phase-space of the system, as defined by the associated \emph{canonical constraints}, i.e. functions $\{\phi_i\}$ in involution\footnote{The vanishing ideal of the constraints forms a Poisson subalgebra (the constraints are first-class). In our construction the contraints are functions on the space of degree-zero boundary fields.} with respect to the Poisson structure induced by $\Omega^\partial$. It can be seen as a cohomological resolution of the quotient of the vanishing set $C\coloneqq\{\phi_i = 0\}$ with respect to the action of symmetries, in the sense that the space of invariant functions on the locus defined by $C$ is the degree-zero cohomology of the BFV complex \cite{Schaetz} $\mathfrak{BFV}^\bullet\coloneqq \left(C^\infty(\mathcal{F}^\partial) ,Q^\partial \right)$.

{

When a theory can be given a BV-BFV description, one can discuss its quantisation in this language \cite{CMR2}. The crucial piece of data in that case is the map $\pi\colon \mathcal{F}\to\mathcal{F}^\partial$ and the relations \eqref{BVBFVeqts}, connecting the BV and BFV data together. In this paper we are interested in purely classical considerations (concerning the nature of the asymptotic symmetries arising when boundaries at infinity are taken into account), and could in principle directly work with some given BFV data. 

\begin{remark}\label{Rem:Classicaldegreezerocharges}
In discussing physical symmetries, a useful interpretation of degree-$1$ fields (ghosts) in the BV formalism is as follows. They can be seen as functionals on the space of symmetry generators, whose evaluation tautologically returns the (degree zero) generator itself, {as in formula \eqref{eq:evaluation}}. In particular, the evaluation of (the ghost-linear part of) $S^\partial$ over gauge generators $\Lambda\in C^\infty(M,\mathfrak{g})$ is a degree-zero functional $S^\partial[\Lambda]$, which we interpret as the classical charge (on shell). At least for gauge theories, this is exactly the Maurer--Cartan form on a principal bundle \cite{Bonora1983, Bonora1988}.
\end{remark}

\subsection{Comparison with Noether procedure}\label{s:noethercomparison}
In this section we show how one recovers the standard Noether analysis of surface charges from the (BV-)BFV picture. This comparison can be carried out precisely when the field theory enjoys symmetries that are closed off-shell. A theory of this kind will be called ``of BRST type'' and for example QED, Yang-Mills and the scalar field theory treated in this work are of this type. For such theories, the BV-BFV data does indeed include a straightforward genearlisation of Noether analysis. The main goal of this section is to prove that, if we denote by $\mathcal{Q}_N[\Lambda]$ the Noether charge associated to a (local) symmetry generated by a gauge parameter $\Lambda$, then we have
\begin{equation}\label{NoetherBVBFV}
    S^\partial[\Lambda]= \mathcal{Q}_N[\Lambda] + \text{higher antifield number}.
\end{equation}

\begin{remark}
Although in the remainder of the paper we use local functionals and local forms on the space of fields, { which typically arise by integrating functional- or form-valued densities\footnote{More precisely, distributions.} over a spacetime manifold, we shall phrase this section in terms of these densities themselves}. This is done in order to make contact with the literature in this field (see e.g. \cite{IW} and \cite{OS} for a more recent discussion). The correspondence between the two becomes one-to-one by correctly considering boundary terms, which we do by means of the map $\pi$ between bulk and boundary fields (cf. Equation \eqref{BVBFVmap}). Then, an expression involving density-valued local forms like $f + dg$ is written as $F + \pi^* G$ where $F=\int_M f$ and $G=\int_{\partial M} g$. For a densitised version of the BV-BFV construction on stratified manifolds see \cite{MSW}.
\end{remark}

In order to compare Noether's analysis with the BV-BFV construction, we use a density version of Equations \eqref{BVBFVeqts}, namely\footnote{The comparison is made by setting $S=\int L$ together with $\Omega=\int \varpi$ and $\alpha=\int \theta$, see Equation \eqref{BVform}. Notice that we require $[Q,d]=0$.}:
\begin{subequations}\label{densityBVBFV}\begin{align}
    \iota_Q\varpi &= \delta L + d\theta^\partial \label{BVeqt1dens}\\
    \frac12\iota_Q\iota_Q\varpi &= d L^\partial \label{mCMEdens}
\end{align}\end{subequations}
and similarly for their boundary counterparts, denoted by the superscript ${}^\partial$. It is also useful to define the BV-BFV difference (see \cite{MSW})
\begin{equation}\label{BVBFVDiff}
    \mathbb{\Delta}^\partial\coloneqq L^\partial - \iota_{Q}\theta^\partial,
\end{equation} 
so that, combining Equations \eqref{BVeqt1dens} and \eqref{mCMEdens}, we can write the failure of the invariance of the BV Lagrangian density $L$ under a gauge transformation to be
\begin{equation}\label{deltaequation}
    \mathcal{L}_Q L = d\left(2L^\partial - \iota_{Q}\theta^\partial\right) = d\left( L^\partial + \mathbb{\Delta}^\partial\right).
\end{equation}

Let us assume that the theory we are interested in is ``of BRST type'', i.e. it is described by a Lagrangian density $L_{\text{cl}}$ on a space of fields $\mathcal{E}_{\BRST}$ that enjoys (off-shell) symmetries encoded in a BRST operator $Q_{\BRST}$, an odd vector field of degree $1$ on $\mathcal{E}_{\BRST}$ such that $[Q_{\BRST},Q_{\BRST}]=0$. Let us denote fields on $\mathcal{E}_{\BRST}$ by $\Phi$ (classical fields and ghosts), in non-negative degrees. If $\theta_N$ is the one-form on $\mathcal{E}_{\BRST}$ obtained by variation of the classical action and integration by parts\footnote{$\theta_N$ is often referred to as the presymplectic potential of the theory.} 
\begin{equation}\label{Noetherform}
    \delta L_{\text{cl}} = \mathbb{EL} + d\theta_N,
\end{equation}
we have
\begin{equation}\label{Noethereq}
    \mathcal{L}_{Q_{\BRST}} L_{\text{cl}} 
    = 
    \iota_{Q_{\BRST}}\mathbb{EL} + d(\iota_{{Q}_{\BRST}}\theta_N) 
    = 
    d B
\end{equation}
for some density $B$, and $\mathbb{EL}$ is a { density valued in} one-forms on $\mathcal{E}_{\BRST}$ that defines the critical locus of the theory (Euler--Lagrange equations of motion).

\begin{remark}
Notice that Equation \eqref{Noethereq} is sometimes expressed in the literature as 
$$
\delta_\Lambda L_{\text{cl}} = \mathbb{EL}(\delta_\Lambda) + d\theta_N(\delta_\Lambda)
$$
where $\delta_\Lambda$ denotes a gauge transformation with gauge parameter $\Lambda$, seen as a vector field on the space of fields. Recall that the ghost fields that appear in the expressions of $Q_{\BRST}$ can be straightforwardly thought of as evaluation functionals on gauge parameters $\Lambda$ (see Remark \ref{Rem:Classicaldegreezerocharges}). The two expressions coincide after evaluation.
\end{remark}

The Noether charge density is defined by 
\begin{equation}\label{NoetherChargeDensity}
   q_N[\Lambda]\coloneqq B[\Lambda] - (\iota_{{Q}_{\BRST}}\theta_N) [\Lambda]
\end{equation}
and it is closed on-shell (for every $\Lambda$), i.e. $dq_N\approx 0$. The Noether charge is then given by integration:
\begin{equation}\label{NoetherCharge}
    \mathcal{Q}_N[\Lambda] := \intl{\partial M} q_N [\Lambda].
\end{equation}

In this scenario, we can implement the following simplifying assumptions on the BV data\footnote{It is important to observe here that cases such as Chern--Simons theory, although they admit a symmetry distribution which is closed off shell, are not always presented as a BRST-type theory in the sense used here, for example when phrased in the AKSZ language. See \cite[Section 1.3 and Proposition 47]{MSW} for more details.}. The space of BV fields will be given by $\mathcal{F}=T^*[-1]\mathcal{E}_{\BRST}$, and we denote antifields by $\Phi^\ddag$.  
{ Let us denote the symplectic density by $\varpi = \delta \theta$, with $\theta(x)=\Phi_x^\ddag \delta\Phi_x$ (cf. with \eqref{BVform}), where $\delta$ is the de Rham differential on $\Fcal$}, and let $\check{Q}_{\BRST}$ be the cotangent lift of $Q_{\BRST}$ to $T^*[-1]\mathcal{E}_{\BRST}$

The BV action functional, in this case, is simply given by 
$$L= L_{\text{cl}} + \iota_{\check{Q}_{\BRST}}\theta.$$ 
The BV operator $Q$ is obtained from the the standard Koszul--Tate--Chevalley--Eilenberg construction for a gauge theory, and coincides with the Hamiltonian vector field of $L$ with respect to the (graded) symplectic form $\int_{M}\varpi$. In the case of a BRST-type theory, $Q$ splits as
$$Q = \check{Q}_{\BRST} + {Q}_{K},$$ 
with ${Q}_{K}$ the Koszul differential\footnote{The ``Tate'' part of the Koszul--Tate differential is encoded in the cotangent directions of $\check{Q}_{\BRST}$.} (only acting on antifields). We recall that $Q_K\Phi^\ddag = \mathbb{EL}^\Phi$ (the $\Phi$ component of $\mathbb{EL})$ and zero otherwise, by definition of the Koszul differential for the critical locus of the classical action $S_{\text{cl}}$. Hence, $\langle Q_K\Phi^\ddag,\delta\Phi\rangle = \mathbb{EL}$.

In order to prove \eqref{NoetherBVBFV} we need to show that 
$$
[L^\partial]_{dR} = [q_N]_{dR} + \text{higher antifield number},
$$
where $[\cdot]_{dR}$ denotes de Rham cohomology classes, and to do this we will first need to show that 
\begin{equation}\label{BVnoetherform}
[\theta^\partial]_{dR} = - [\theta_N]_{dR} + \text{higher antifield number}.
\end{equation}
To this end, we observe that
\begin{align}
d\theta^\partial &= \iota_Q\varpi^\partial - \delta L 
    = 
    \langle Q{\Phi^\ddag},\delta\Phi \rangle 
        + \langle \delta\Phi^\ddag, Q\Phi\rangle - \delta L \\
    & = \mathbb{EL} + \langle \check{Q}_{\BRST}{\Phi^\ddag}, \delta\Phi\rangle + \langle \delta\Phi^\ddag, Q\Phi\rangle - \delta L_{\text{cl}} - \delta(\iota_{\check{Q}_{\BRST}}\theta)\\
    & = -d\theta_N + \langle \check{Q}_{\BRST}{\Phi^\ddag}, \delta\Phi\rangle + \langle \delta\Phi^\ddag, Q\Phi\rangle - \delta(\iota_{\check{Q}_{\BRST}}\theta),
\end{align}
where we have used the splitting of $Q=\check{Q}_{\BRST} + Q_K$, and the explicit formula for $L$ in the second line, and \eqref{Noetherform} to get the third line. Hence, we prove equation \eqref{BVnoetherform} if we can show that $d\theta^\partial + d\theta_N$ is at least linear in antifields. But it clearly is, since
$$
d\theta^\partial + d\theta_N = \langle \check{Q}_{\BRST}{\Phi^\ddag}, \delta\Phi\rangle + \langle \delta\Phi^\ddag, Q\Phi\rangle - \delta(\iota_{\check{Q}_{\BRST}}\theta) =\langle \check{Q}_{\BRST}{\Phi^\ddag}, \delta\Phi\rangle + \langle \Phi^\ddag,\delta (Q\Phi)\rangle,
$$
and $\check{Q}_{\BRST}{\Phi^\ddag}$ is necessarily at least linear in ${\Phi^\ddag}$. This implies that
\begin{equation}\label{higherremainder}
    d(\iota_Q \theta^\partial + \iota_Q\theta_N) = d(\iota_Q \theta^\partial + \iota_{{Q}_{\BRST}}\theta_N) = \iota_Q\left(  \langle \check{Q}_{\BRST}{\Phi^\ddag}, \delta\Phi\rangle + \langle \Phi^\ddag,\delta (Q\Phi\rangle)\right) 
\end{equation}
is also higher in antifield number.

Now, with a little work one can check that $\mathbb{\Delta}^\partial=B$ (we refer to \cite[Theorem 31]{MSW}), so that, from Equation \eqref{BVBFVDiff} we can argue that
\begin{align*}
    dL^\partial &= d\left(\mathbb{\Delta}^\partial + \iota_{Q}\theta^\partial\right) \\
    & = d\left( B - \iota_{\check{Q}_{\BRST}}\theta_N  + \text{higher antifield number}\right) \\
    & = d\left( q_N + \text{higher antifield number} \right),
\end{align*}
where we have used Equation \eqref{higherremainder} and $\Delta^\partial = B$ in the second line, so that the boundary BFV action reads:
\begin{equation}
    S^\partial = \intl{\partial M} q_N +  \text{higher antifield number}.
\end{equation}
Evaluating on a gauge parameter we obtain
\begin{equation}
    S^\partial[\Gamma] = \mathcal{Q}_N[\Gamma] +  \text{higher antifield number},
\end{equation}
as claimed.

The information contained in $S^\partial $ is at least twofold. It generates gauge transformations \emph{via} its Hamiltonian vector field (the BFV operator $Q^\partial$) and, as we have seen, it computes the Noether charge. Simultaneously, one recovers the canonical constraints by treating ghost fields in $S^\partial$ as Lagrange multipliers\footnote{For the application of this point of view to the nontrivial cases of General Relativity in the Einstein--Hilbert and Palatini--Cartan formalisms see \cite{CSEH,CCS}.}. 
As such, it provides a straightforward generalisation of Noether charges in the case where the symmetries of the theory do not close off shell, and it also contains dynamical information.

\begin{remark}\label{rem:holography}
In the above construction we have seen how the failure of gauge invariance of the classical Lagrangian, the boundary term $B$, is controlled --- in the BV-BFV setting --- by the ``BV-BFV difference'' $\mathbb{\Delta}^\partial\coloneqq L^\partial - \iota_{Q}\theta^\partial$. This functional was defined and discussed in detail in \cite[Definition 21]{MSW}, and it was shown to encode the failure of gauge invariance of classical data. Often one finds that $\mathbb{\Delta}^\partial=0$, but this is not always the case. This observation is linked to gauge anomalies, descent equations and holography. While the example of Electrodynamics considered in this paper is such that $\mathbb{\Delta}^\partial=0$, the sourced dual model of Section \ref{Sec:hardscalar} does not, similarly to theories like Chern--Simons or $BF$ theory in dimension 3 or higher. This feature does not seem to impact the description of asymptotic symmetries, and further investigation on the consequences of this observation will be deferred to future work.
\end{remark}

\subsection{Extension to corners}\label{Corners}
We would like to discuss the extension of the BV-BFV relations of Equations \eqref{BVBFVeqts} to higher codimension strata like corners. This point of view was presented systematically in \cite{CMR1}. A density version of this construction, and its relation to holography, is given in \cite{MSW}.

When the boundary of a manifold $M$ has a boundary of its own (a codimension $2$ stratum for $M$), Equation \eqref{boundaryBVBFV} and \eqref{boundaryCME} will typically no longer be satisfied. In good cases, one can associate additional cohomological structure to corners so that equations analogous to \eqref{BVBFVeqts} are satisfied. With an abuse of notation, let us denote said data by $(\mathcal{F}^{\partial\partial},\Omega^{\partial\partial},S^{\partial\partial},Q^{\partial\partial})$, where $(\mathcal{F}^{\partial\partial},\Omega^{\partial\partial})$ is a $1$-symplectic manifold associated to the corner\footnote{Typically this turns out to be the restriction of fields to the corner, possibly with some additional reduction.}, then:
\begin{subequations}
    \begin{align}\label{cornerBVBFV}
    \iota_{Q^\partial }\Omega^\partial  
        &= \delta S^\partial + \pi^*_\partial\alpha^{\partial\partial}\\\label{cornerCME}
    \frac12 \iota_{Q^\partial }
        \iota_{Q^\partial }\Omega^\partial  
        &= \pi^*_\partial S^{\partial\partial}
    \end{align}
\end{subequations}
where $\pi_\partial \colon \mathcal{F}^\partial \to \mathcal{F}^{\partial\partial}$ is a surjective submersion connecting the boundary and corner data. Observe that $\Omega$ and $\Omega^\partial$ are linked by\footnote{The density version of this equation is $\mathcal{L}_Q\varpi = d\varpi^\partial$.}
\begin{equation}
    \mathcal{L}_Q \Omega = \pi^*\Omega^\partial
\end{equation}
as can be checked by applying $\delta$ to Equation \eqref{BVBFV}.  Analogously, as a consequence of \eqref{cornerBVBFV}, we have
\begin{equation}
    \mathcal{L}_{Q^\partial} \omega^\partial = \pi_\partial^*\Omega^{\partial\partial}.
\end{equation}

The underlying philosophy, here, is that out of the tower of BV-BFV relations one can extract an inhomogeneous local form valued densities $\mathcal{O}^\bullet\in\Omega^{\bullet,\bullet}_{\text{loc}}(M,\mathcal{F})$ which satisfy the descent equation \cite{MSZ,ZUMINO1985477}
\begin{equation}\label{descent}
    (\mathcal{L}_Q - d) \mathcal{O}^\bullet=0.
\end{equation}
For example, one can construct $\varpi^\bullet\in\Omega^{\bullet,2}_{\text{loc}}(M, \mathcal{F})$ such that (denote by $K$ the corner of $M$)
\begin{equation}\label{formsfromdensity}
    \Omega=\intl{M} \varpi^\bullet; \qquad \Omega^\partial = \intl{\partial M} \varpi^\bullet; \qquad \Omega^{\partial\partial}=\intl{K}\varpi^{\bullet}; \qquad \dots
\end{equation}
and so on. Then, Equation \eqref{descent} encodes the appropriate BV-BFV relations between strata of codimension $k$ and $k+1$.

\begin{remark}
Observe that, in principle, one can make the BV Lagrangian $L$ into a solution $L^\bullet$ of the descent equations as well. A universal solution was presented in \cite[Theorem 23]{MSW}. Notice that higher codimension data controls the failure of gauge invariance at lower codimensions.
\end{remark}

\subsection{Extended symmetries}\label{extgauge}
When discussing gauge field theory, one typically requires the gauge parameters generating a gauge transformation to be compactly supported in the bulk manifold. A \textit{symmetry}, then, is supposed to be a transformation that not only preserves the action functional of the theory, but also the canonical symplectic form.

When said compact support is not required of the gauge parameters, it is often the case that the quantities above will fail to be invariant, due to emerging boundary or corner terms. However, according to the philosophy presented in Section \ref{Corners}, the non-invariance of a particular piece of data is not relevant per se, as long as it can be controlled (or compensated). Observe, indeed, that the possible failure of gauge invariance of the Lagrangian --- encoded in the boundary term $B$ of equation \eqref{Noethereq} --- plays a fundamental in computing Noether charges (Equations \eqref{NoetherChargeDensity} and \eqref{NoetherCharge}).

In this spirit, one can extend the notion of \emph{symmetry} of the theory, regardless of whether field transformations also preserve the (degree-zero) symplectic structure canonically associated to the theory. Indeed, as long as one keeps track of the failure of gauge invariance of the relevant data at every codimension, it is possible to recover invariance as a whole in terms of composite objects that satisfy the descent equations \eqref{descent}.

\begin{remark}
The perspective outlined in this section will become important when discussing the interpretational coundrum of whether asymptotic charges should be thought of as generators of large gauge transformations, or not. They will be interpreted as such, in this extended sense, in Section \ref{EMcorners}.
\end{remark}

}

\subsection{Geometric conventions}\label{geometricpreliminaries}
When not stated otherwise, in this paper we work with the Minkowski spacetime with signature $(1, -1,-1,-1)$. However, our constructions can be adapted to the case of asymptotically-flat spacetimes.

For our construction of asymptotic charges, we begin with identifying a sufficiently large, precompact region $\Wcal_R$ inside our space-time, bounded by a piecewise-null and piecewise spacelike boundary $\partial\Wcal_R := \scri^+_r \cup \scri^-_r \cup \Hcal^+_\tau \cup \Hcal^-_\tau$, with $R>0$, as shown in Figure \ref{diagram}. Later on we will take a limit, where this region is enlarged ``to infinity''.

\begin{figure}[htb]
\centering
\begin{tikzpicture}[x=1.00mm, y=1.00mm, inner xsep=0pt, inner ysep=0pt, outer xsep=0pt, outer ysep=0pt]
\path[line width=0mm] (58.13,67.94) rectangle +(84.61,64.04);
\definecolor{L}{rgb}{0,0,0}
\definecolor{F}{rgb}{0.753,0.753,0.753}
\path[line width=0.30mm, draw=L, fill=F] (70.13,100.23) .. controls (70.13,100.23) and (70.13,100.23) .. (70.13,100.23) .. controls (75.08,105.19) and (84.96,115.06) .. (89.99,120.10) .. controls (90.03,120.14) and (90.06,120.24) .. (90.11,120.22) .. controls (94.18,118.46) and (106.23,118.30) .. (109.58,120.12) .. controls (109.70,120.19) and (109.71,120.32) .. (109.85,120.32) .. controls (114.92,115.32) and (125.18,105.14) .. (130.29,100.11) .. controls (130.31,100.09) and (130.34,100.07) .. (130.35,100.05) .. controls (132.30,94.94) and (68.56,94.51) .. (70.13,100.23) -- cycle;
\path[line width=0.30mm, draw=L, dash pattern=on 2.00mm off 1.00mm] (100.10,99.88) [rotate around={360:(100.10,99.88)}] ellipse (30.18mm and 3.67mm);
\definecolor{L}{rgb}{1,0,0}
\path[line width=0.30mm, draw=L] (110.16,79.88) -- (130.63,99.83);
\path[line width=0.30mm, draw=L] (69.88,99.94) -- (90.19,79.88);
\definecolor{L}{rgb}{0,0,0}
\path[line width=0.30mm, draw=L] (90.09,80.28) .. controls (93.52,82.99) and (95.56,84.22) .. (100.53,84.16) .. controls (105.29,84.10) and (107.80,82.60) .. (110.16,80.22);
\path[line width=0.30mm, draw=L] (60.13,100.01) -- (140.73,99.93);
\definecolor{F}{rgb}{0,0,0}
\path[line width=0.30mm, draw=L, fill=F] (140.73,99.93) -- (139.33,100.63) -- (139.33,99.23) -- (140.73,99.93) -- cycle;
\path[line width=0.30mm, draw=L] (100.03,70.29) -- (100.03,128.48);
\path[line width=0.30mm, draw=L, fill=F] (100.03,128.48) -- (99.33,127.08) -- (100.73,127.08) -- (100.03,128.48) -- cycle;
\path[line width=0.30mm, draw=L, dash pattern=on 0.30mm off 0.50mm] (140.31,79.13);
\path[line width=0.30mm, draw=L, dash pattern=on 0.30mm off 0.50mm] (70.14,69.94) -- (129.90,129.89);
\path[line width=0.30mm, draw=L, dash pattern=on 0.30mm off 0.50mm] (69.95,129.98) -- (129.90,70.03);
\path[line width=0.30mm, draw=L, dash pattern=on 0.30mm off 0.50mm] (76.05,109.34) arc (0:0:0.00mm) -- cycle;
\path[line width=0.30mm, draw=L] (90.09,120.16) .. controls (93.94,117.81) and (97.06,116.45) .. (100.11,116.28) .. controls (102.50,116.15) and (106.34,117.94) .. (110.15,120.22);
\definecolor{L}{rgb}{1,0,0}
\path[line width=0.30mm, draw=L] (100.09,120.32) [rotate around={360:(100.09,120.32)}] ellipse (9.85mm and 1.25mm);
\path[line width=0.30mm, draw=L] (100.23,80.12) ellipse (9.99mm and 1.49mm);
\path[line width=0.30mm, draw=L] (69.77,99.88) -- (90.17,120.37);
\path[line width=0.30mm, draw=L] (109.93,120.30) -- (130.64,99.78);
\node at (104,121) {$\Hcal_\tau^+$};
\node at (104,79) {$\Hcal_\tau^-$};
\node at (115,90) {$\scri_r^-$};
\node at (115,110) {$\scri_r^+$};
\end{tikzpicture}%
\caption{Region $\Wcal_R$ inside our space-time, bounded by a piecewise-null and piecewise spacelike boundary $\partial\Wcal_R := \scri^+_r \cup \scri^-_r \cup \Hcal^+_\tau \cup \Hcal^-_\tau$\label{diagram}.}
\end{figure}

In this work we will use two ways of parametrising the boundary: the $R,s,\bl$ variables of Herdegen (see e.g. \cite{Her95,Her16}) and the retarded light-cone coordinates. We present our results in both parametrisations, not only to make it easier to understand for different communities, but also because techniques used in proofs of our main results are slightly different and it is instructive to see both.

\subsubsection{$R,s,\bl$ variables}\label{sec:Rsl}
One way to describe null asymptotics of fields is to use a set of variables introduced by Herdegen, which we refer to as $R,s,\bl$ variables in this work. Let $\bl$ is a future-pointing null vector, $\bt$ a future-pointing timelike vector\footnote{In \cite{Her16} the author uses $l$ and $t$ rather than $\bl$ and $\bt$, but we want to avoid confusion with the notation for the time coordinate.} and $R,s\in \RR$, $R\geq 0$. { Note that this set of parameters seems over-complete. There are two scalars and one null vector, so altogether 5 free parameters. This redundancy is not a problem, since $\bl$ runs over null directions, rather than null vectors.}

In \cite{Her16} (and previous works), one uses these parameters to define a space-time point $x$ according to:
\[
x=R\bl + s\frac{\bt}{\bt\cdot \bl}\,,
\]
More about variables $R,s,\bl$ can be found in \cite[Appendix B]{Her16}. In particular, a differentiable field $B$ on Minkowski spacetime, defines $\beta(R,s,\bl)=B(x)$ with the scaling property: $b(R/\lambda, \lambda s, \lambda \bl)=b(R,s,\bl)$, $\lambda>0$. Denote
\[
L_{ab}=\bl_a \frac{\partial}{\partial \bl^b}-\bl_b \frac{\partial}{\partial \bl^a}\,.
\]
One can show that:
\be\label{Rsl}
\frac{\partial}{\partial x^b} B(x)=\bl_b \dot{\beta}(R,s,\bl)+\frac{\bt^a}{R \bt \cdot \bl} L_{ab} \beta(R,s,\bl)\,,
\ee
where the dot denotes the derivative with respect to $s$ and
\[
L_{ab} \beta(R,s,\bl)=R\left(\bl_a \frac{\partial}{\partial x^b}  B(x)-\bl_b \frac{\partial}{\partial x^a}  B(x)\right)\,
\]
Very often we will use integration over the set of null directions. Let
\[
C_+\doteq \{\bl|\bl\cdot \bl =0,\bl^0>0\}\,.
\]
and for a fixed $\bt$, define $C_+^{\bt}$ as the intersection of $C_+$ with the $\bt\cdot\bl=0$ plane. $C_+^{\bt}$ is a unit sphere in this plane and hence can be equipped with the usual metric whose line elements is denoted by $d\Omega^2$. Following \cite{Her16}, let $f(\bl)$ be a measurable function on $C_+$, homogeneous of degree $-2$. The integral defined by
\[
\int f(\bl)d^2\bl\doteq \int_{C_+^{\bt}} f(\bl) d\Omega^2
\]
does not depend on the choice of the vector $\bt$. In the present paper, we will often make use of this fact and identify the integral over the null directions on the left-hand side with the integral over a concrete unit sphere determined by the choice of $\bt$. In a fixed coordinate system, the natural choice is: $\bt=(1,0,0,0)$. We come back to this at the end of the next section.
\subsubsection{Retarded light-cone coordinates}\label{Sect:coordinates}
Another convenient way to describe the null asymptotic of smooth fields on Minkowski spacetime makes use of \textit{retarded coordinates}. We start with the standard set of coordinates $(t,x,y,z)$ and introduce space-like spherical coordinates $(r,x^A)$, with $x^A$, $A=1,2$. Next, we define \emph{retarded} (resp. advanced) light-cone coordinates as  $u_\pm=t \mp r$. 

In coordinates $(u_\pm,r,x^1,x^2)$, a line element in Minkowski metric reads:
\begin{equation}\label{retardedlineelement}
    ds_\pm^2= +du_\pm^2 \pm 2drdu_\pm - r^2d\Omega^2\,,
\end{equation}
where $d\Omega^2$ is the line element for the unit 2-sphere\footnote{In \cite{Strominger} the unit line element on $S^2$ is expressed in complex coordinates as $d\Omega^2=2\gamma_{z\bar{z}}dzd\bar{z}$, with $\gamma_{z\bar{z}}=(1+z\bar{z})^{-2}$, while in \cite{Cam15,CL15,CE17} $d\Omega^2=q_{AB}dx^Adx^B$.}. The matrix representation of Minkowski metric is
\begin{equation}\label{retardedmink}
	g_\pm = \left(\begin{array}{cc}
		\begin{array}{cc}+1 & \pm 1 \\
		\pm 1 & 0 
		\end{array} & \mathbb{0}\\
		\mathbb{0} &  -r^2g_{S^2}
	\end{array}\right),
\end{equation}
with determinant $\mathrm{det}(g_\pm) = -r^4 \det(g_{S^2})$, and the inverse 
\begin{equation}\label{inverseretardedmink}
	g_\pm^{-1} = \left(\begin{array}{cc}
		\begin{array}{cc} 0 & \pm 1 \\
		\pm 1 & -1 
		\end{array} & \mathbb{0}\\
		\mathbb{0} &  -r^{-2}g^{-1}_{S^2}
	\end{array}\right).
\end{equation}

Let $\{x^A\}_{A=1,2}$ be coordinates of a point on the unit two-sphere \footnote{We refer to points on the unit two-sphere as $x^A$, using the abstract index notation.} embedded in  three-dimensional Euclidean space. We denote the corresponding point of this three-dimensional space by $\hat{x}$. In this parametrisation, a spacetime point can be written as:
\[
 	x=\left(\begin{array}{c}
 	u_++r\\
 	r \hat{x}
 	\end{array}\right)\,.
\]
To relate this particular coordinatisation to the formulation using $(R,s,\bl)$ variables, choose $\bt=(1,0,0,0)$ and consider null vectors of the form $\bl=(1,\hat{x})$, where $\hat{x}$ is the unit three-vector determined by a sphere point $x^A$. Then identify $R$ with the radial coordinate $r$, so that:
\[
x=r \left(\begin{array}{c}
1\\
\hat{x}
\end{array}\right) + s \left(\begin{array}{c}
1\\
0
\end{array}\right)=r \left(\begin{array}{c}
1\\
\hat{x}
\end{array}\right) + u_+ \left(\begin{array}{c}
1\\
0
\end{array}\right)
\]
where we used the fact that $\bl\cdot \bt=1$ and we identified Herdegen's variable $s$ with the retarded time $u_+$, since $s+r=t$ for our choice of $\bl$ and $\bt$. Observe that, in particular, in the $(r,u_+)$ coordinates we have $\bl^r=1$ and $\bl^{u_+}=0$.

\subsection{Parametrisation of the boundary}
In this paper we are concerned with symmetries and associated charges that appear on asymptotic boundary components and corners (i.e. boundaries of boundary components). For example, we will consider surfaces at constant coordinate \emph{radius} $R$ and then take the limit for $R\to\infty$  (see Figure \ref{diagram}).

We denote by $\mathcal{I}^{\pm}$ the copy of $S^2\times \mathbb{R}$ obtained after taking the $r\to +\infty$ (or $R\to +\infty$ in the other description) limit while keeping $u_\pm$ constant. We treat the limits $\mathcal{I}^{\pm}$ as boundaries at infinity, and call them \emph{future/ past null infinity}. From $\mathcal{I}^+$, one gets two connected components of $\partial \mathcal{I}^+$, denoted $\mathcal{I}^+_\pm$ and topologically homeomorphic  to two-dimensional spheres, obtained by taking the limits $u_+\to \pm \infty$, respectively. Similar considerations apply to $\mathcal{I}^-$.

\subsubsection{Hyperbolic coordinates}\label{sec:hyperbolic} Hyperbolic coordinates are used to analyse the behavior of smooth fields in Minkowski spacetime at timelike infinity. They are defined by
 	\[
 	\tau=\sqrt{t^2-r^2}\,,\qquad \rho=\frac{r}{\sqrt{t^2-r^2}}\,,
 	\]
 	so that
 	\[
 	t=\tau\sqrt{1+\rho^2}\,,\qquad r=\rho\tau\,.
 	\]
In these coordinates,  the line element reads:
\begin{equation}
    ds^2 = d\tau^2 - \tau^2\left((1+\rho^2)^{-1} d\rho^2 + \rho^2 d\Omega^2\right),
\end{equation}
and future timelike infinity $i^+$ is obtained by taking the limit $\tau \to \infty$.
A spacetime point can then be written as
 	\[
 	x=\left(\begin{array}{c}
 	\tau\sqrt{1+\rho^2}\\
 	\tau\rho \hat{x}
 	\end{array}\right).
 	\]
Finally, a point on the unit hyperboloid $\Hcal^+$ (i.e. $\tau=1$) is written as
 	\[
 	Y^\mu=(\sqrt{1+\rho^2},\rho\hat{x})\,.
 	\]

\subsection{Differential forms conventions}
A differential $k$-form is written in a local coordinate chart as
$$
    \alpha = \frac{1}{k!} \alpha_{\mu_1\dots \mu_k} dx^{\mu_1}\wedge \dots \wedge dx^{\mu_k}
$$
with $\alpha_{\mu_1\dots\mu_k}$ totally antisymmetric in the indices.
The operation of taking the Hodge dual on a generating set of $k$-forms over an $N$-dimensional (pseudo)-Riemannian manifold $(M,\eta)$ is given by
$$(\star dx^{\nu_1} \dots dx^{\nu_k}) = \frac{\sqrt{|g|}}{(n-k)!} \eta^{\nu_{1} \rho_{1}} \dots \eta^{\nu_{k} \rho_{k}} \epsilon_{\rho_1\dots \rho_k\mu_{k+1} \dots \mu_{N-k}}  dx^{\mu_{k+1}} \dots dx^{\mu_{N}},$$
which, for $\alpha,\beta\in\Omega^k(M)$ yields
$$
    \alpha\wedge \star \beta = \frac{1}{k!}\alpha_{\mu_1\dots \mu_k}\beta^{\mu_1\dots \mu_k}\mathrm{dVol}_\eta\,.
$$
The indices are raised with the inverse metric, e.g. $\beta^\mu = g^{\mu\nu}X_\nu$ denotes the components of the vector $\beta=(g^\flat)^{-1}(X)$, for $g^\flat\colon TM \stackrel{\sim}{\longrightarrow} T^*M$.

The Laplace--Beltrami operator on a Lorentzian manifold of signature $(1,-1,-1,-1)$ is $\Box=-(d d^* + d^* d)$, with the codifferential defined by
$$d^*\equiv\star d\star\colon \Omega^k(M)\to\Omega^{k-1}(M).$$ 
Its restriction to co-closed forms, i.e. forms in the Lorenz gauge $d^* A = 0$, is $\Box\vert_{\text{coclosed}} = - d^* d$.

\section{Electrodynamics}\label{s:electrodynamics}

In this section we consider electrodynamics,\footnote{Throughout, we use the standard formulation of electrodynamics as a second-order field theory, i.e. where equations of motion are second order. Other literature prefer to employ the first order formulation instead, whose BV-BFV description can be found in \cite{CMR1}.} phrased in the Batalin--Vilkovisky language, to show how we can recover asymptotic symmetries from the appropriate manipulation of the BFV data. 

\subsection{BV-BFV approach to electrodynamics} 
Electrodynamics is formulated in terms of a  $U(1)$ Yang--Mills field theory coupled to matter. For simplicity, we model matter as a complex scalar field, but the same analysis can also be performed for Fermions. For a principal $U(1)$ bundle $P\longrightarrow M$ on a Lorentzian spacetime $(M,g)$, possibly with boundary\footnote{We will later restrict to a situation where  $(M,g)$ is a region in Minkowski spacetime.}, and given an associated $\mathbb{C}^2$ bundle\footnote{On Minkowski spacetime the $\mathbb{C}^2$-bundle is trivial.} $\mathcal{V}\to M$, the extended space of field configurations is 
\begin{equation}
	\mathcal{F}=T^*[-1]\left(\mathcal{A}_P\times \Omega^0(M,\mathcal{V}) \times \Omega^0[1](M)\right)\,,
\end{equation}
where $\mathcal{A}_P$ is the space of electromagnetic potentials { $A$, which we will identify with the space of fluctuations around a reference connection, $\Omega^0(M,\mathcal{V})$ is the space of complex scalar fields $(\ph,\oph)$, and $c\in\Omega^0[1](M)$ denotes ``ghost'' fields.}

There is a canonical shifted symplectic structure $\Omega$ on $\Fcal$, formally given by
$$
    \Omega=\int_M \delta A\delta A^\ddag + \delta c \delta c^\ddag + \delta\ph\delta\ph^\ddagger + \delta \oph \delta \oph^\ddag,
$$
where we denoted fields in the cotangent fiber (also called anti-fields) by $A^\ddag\in \Omega^3[-1](M)$, $\ph^\ddagger\in\Omega^{\text{top}}[-1](M,\mathcal{V}^*)$ and $c^\ddag\in\Omega^{\mathrm{top}}[-2](M)$. Covariant derivatives for the fields $\ph, \oph$ are defined by $d_A \varphi = d \varphi + i q A \varphi$ and $d_A \oph = d \oph - i q A \oph$, with $q\in \mathbb{R}$ a coupling constant. The BV-extended action functional is then given by:
\begin{equation}
	S= \intl{M} \left(-\frac{1}{8\pi} F_A\wedge \star F_A + \frac12 \left(d_A\overline{\ph}\wedge \star d_A\ph + m^2\oph\ph \right) + A^\ddag\wedge d_Ac+\ph^\ddag c\ph - \oph^\ddag c\oph\right)\,,
\end{equation}
where $\star$ is the Hodge operator induced by the Lorentzian structure on $(M,g)$, and the BV operator $Q$ is given by
\begin{eqnarray*}
    QA = d_A c & QA^\ddag = -\frac{1}{4\pi}d_A\star F_A - iq\,\oph\star d_A\ph + iq\,\star d_A\oph \ph \\
    Q\ph= c\ph & Q\ph^\ddagger= (-d_A\star d_A + m^2)\oph + \ph^\ddag c \\
    Q\oph= -c\oph & Q\oph^\ddagger= (-d_A\star d_A + m^2)\ph - \oph^\ddag c\\
        Qc=0 & Qc^\ddag =0
\end{eqnarray*}
where we used that $d_A\overline{\ph}\wedge \star d_A\ph= - \star d_A\overline{\ph}\wedge  d_A\ph$.

Denote the matter current by $J:= -iq\,\oph\star d_A\ph + iq\,\star d_A \oph\ph$.  The classical equations of motion (the degree-zero sector of the condition $Q=0$) are given by:
\begin{subequations}\label{EOM}\begin{align}
    d_A\star F_A &= J\\
    (-d_A\star d_A + m^2)\ph &=0\\
    (-d_A\star d_A + m^2)\oph &=0
\end{align}\end{subequations}
Note that $d_A J=0$ on shell, since $[F_A,\star F_A]=0$.

Since  Yang--Mills theory satisfies the 
BV-BFV axioms stated in equation \eqref{BVBFVeqts} (see e.g. \cite{CMR1,MSW} for details), we obtain --- on a manifold with boundary --- the following BFV data $(\mathcal{F}^\partial, S^\partial, Q^\partial, \Omega^\partial)$:
\begin{itemize}
	\item The \emph{space of boundary fields} is
	$$\mathcal{F}^\partial\coloneqq T^*\left(\mathcal{A}_{\iota^*P}\times \Omega^0(\partial M,\mathcal{V})\times \Omega^0[1](\partial M)\right)\,,$$
	where we denoted by $\mathcal{A}_{\iota^* P}$ the space of connections\footnote{Since in the bulk we considered fluctuations around a reference connection, the fields in $\mathcal{A}_{\iota^*P}$ can also be thought of as fluctuations.} on the induced principal bundle $\iota^*P$ on $\partial M$.
\item The \emph{boundary action} is given by
\begin{equation}\label{EMFullBoundaryaction}
    S^{\partial}=\frac12 \iota_Q\iota_Q\Omega
    = 
    \int_{M} d\left[ c  \left(-\frac{1}{4\pi}d_A\star F_A  + J\right)\right]\,,
\end{equation}
which is a degree $1$ functional on $\mathcal{F}^\partial$.
\item $\mathcal{F}^\partial$ is equipped
with the symplectic form $\Omega^\partial$ given by
$$
    \Omega^\partial = \intl{\partial M} \frac{1}{4\pi}\delta A \delta [\star F_A]_{\partial M} + \delta c\delta A^\ddag + \delta\oph \delta[\star d_A \ph]_{\partial M} + \delta \ph \delta[\star d_A \oph]_{\partial M},
$$
\item {The BFV operator $Q^\partial$ is the Hamiltonian vector field of $S^\partial$, i.e. 
$$\iota_{Q^\partial} \Omega^\partial = \delta S^\partial.$$
}
\end{itemize}

The projection map $\pi\colon \mathcal{F}\to \mathcal{F}^\partial$ is simply the restriction of fields and normal jets to the boundary, composed with a redefinition of fields sending a normal jet of $A$ (resp $\varphi$) - restricted to the boundary - to $[\star F_A]_{\partial M}$ (resp. $[\star d_A\varphi]_{\partial M}$), which is considered an independent field. A careful analysis of the symplectic manifold of boundary fields for the scalar case was given in \cite{CattaneoMnev}.

\begin{remark}\label{rem:cornerterms}
To obtain \eqref{EMFullBoundaryaction} we could make the following alternative choice (recall $d_Ac$ has even parity {, since $c$ is odd}): 
$$
\int_M d_A c\wedge d_A \star F_A  = \int_M d(c d_A\star F_A) = \int_M d(d_A c \wedge\star F_A )
$$ 
{ Observe that this choice yields the same result without generating corner terms}, but is better suited to recover Herdegen's formulas for soft charges \cite{Her16}, while  \eqref{EMFullBoundaryaction}, which has the advantage of manifestly vanishing on shell, will be useful to reproduce formulas in \cite{Strominger}.
\end{remark}

 More on the BV-BFV structure of Yang-Mills theory in the second order formalism and its relation with extended phase spaces and edge modes can be found in \cite{MSW}, while the first order formulation has been discussed in \cite{CMR1}.

\begin{remark}
Note that the BV data presented above has been historically associated to fields on compact manifolds, or equipped with vanishing boundary conditions, or given for compactly supported fields. As such, it was never set up to interact with a choice of fall-off conditions on fields. However, said conditions can be introduced once one extends the BV-BFV construction to noncompact manifolds, by defining fields to be sections of bundles supplemented with the appropriate falloff conditions. This is what we will do in the next section.
\end{remark}

\subsection{Asymptotic fields}
In this work, instead of considering $S^\partial$ at a finite boundary, we consider  $S^\partial$ at infinity. To make this precise, we need to impose some fall-off conditions on the variables $A$ and $\ph$, to ensure that their asymptotes are well defined. Note that the natural limit for the electromagnetic potential is the null infinity $\scri=\scri^+\cup \scri^-$, while the matter current $J$ propagates to time-like infinity $\Hcal=\Hcal^+\cup\Hcal^-$. This translates to the requirement that
\be\label{Eq:hyperboloidfall-off}
\lim_{\tau\to \infty} A\vert_{\Hcal^\pm_\tau} = \lim_{r\to R} J\vert_{\scri^\pm}=0.
\ee

\subsubsection{Free electromagnetic field}
In the notation of \cite{Her95,r05903,r05892,r05873,Her16}, the asymptotic electromagnetic potentials are defined as follows:
\begin{align*}
\lim\limits_{R\rightarrow \infty} RA(x+R\bl)&=V(x\cdot \bl,\bl)\,,\\
\lim\limits_{R\rightarrow \infty} RA(x-R\bl)&=V'(x\cdot \bl,\bl)\,.
\end{align*}
Using instead the $(r,u_+,z,\bar{z})$ coordinates, one expands $A$ as:
\begin{equation}\label{Afall-off}
    A= \sum_{k=1} \frac{1}{r^k}A^{(k)}\,,
\end{equation}
so that
\[
V(s,\bl)=A^{(1)}(u_+,\hat{x})\,,
\]
where $s=u_+$, $\bl=(1,\hat{x})$ and $\hat{x}$ is a point on unit 2-sphere embedded in 3-dimensional Euclidean space, as explained in section \ref{Sect:coordinates}.

Without external currents ($J=0$) and assuming Lorenz gauge, $A$ satisfies the wave equation
\be\label{wave}
\Box A=0
\ee
and the asymptotic fields have the ``vanishing property'':
\be\label{eq:vanishing}
V(+\infty,\bl)=0=V'(-\infty,\bl)\,,
\ee
i.e. these vanish at time-like infinity. In \cite{Her95,r05903,r05892,r05873,Her16} these also satisfy the following fall-off conditions:
\be\label{Afaloff1}
|V_a(s,\bl)|<\frac{const.}{s^\epsilon}\,,
\ee
\be\label{Afaloff2}
|\dot{V}_a(s,\bl)|<\frac{const.}{s^{1+\epsilon}}\,,
\ee
similarly for $V'$, but with the role of $-\infty$ and $+\infty$ exchanged.
\subsubsection{Fields with sources}
Now let us consider the equation with sources:
\be\label{eq:source}
\Box A(x)=4\pi J(x)\,.
\ee
For this equation we know that the retarded and advanced Green functions $\Delta^{\rm R/A}$ exist.
We want the current $J$ to describe incoming and outgoing matter fields in a scattering experiment and the free radiation field
\[
A^{\rm rad}=A^{\rm R}-A^{\rm A}
\]
should satisfy the fall-off conditions \eqref{Afaloff1} and \eqref{Afaloff2}. The Pauli-Jordan function is defined by $\Delta=\Delta^{\rm R}-\Delta^{\rm A}$. We have, in relative coordinates,
\[
\Delta(x)=\frac{1}{2\pi}\mathrm{sgn}(x^0)\delta(x^2)\,.
\]
In \cite{Her95}, $\Delta$ is represented as
\[
\Delta(x)=-\frac{1}{8\pi^2} \int \delta'(x\cdot \bl)d^2\bl\,,
\]
so the radiation field
\begin{multline}
    A^{\rm rad}(x)=4\pi \int \Delta(x-y)J(y) dy=
    -\frac{1}{2\pi} \int dy \int d^2\bl\, \delta'((x-y)\cdot \bl) J(y)\\
    =-\frac{1}{2\pi}  \int d^2\bl\int dy(\delta'(x\cdot \bl-y\cdot \bl) J(y))= -\frac{1}{2\pi}  \int \dot V_J(x\cdot \bl,\bl) d^2\bl\,,
\end{multline}
where
\[
V_J(s,\bl)=\int dy \delta(s-y\cdot \bl) J(y)
\]
and we have
\begin{align}
\lim_{R\rightarrow \infty} RA^{\rm R}(x-R\bl)&=V_J(-\infty,\bl)\,,\label{time:ret}\\
\lim_{R\rightarrow \infty} RA^{\rm A}(x+R\bl)&=V_J(+\infty,\bl)\,,\label{time:adv}
\end{align}
from the definition of retarded and advanced solutions.

Assume that $V_J$ is well-defined and that $\dot{V}_J$ satisfies \eqref{Afaloff2}. For physical reasons (see \cite{Her16}), assume that for $v$ on the unit future hyperboloid (i.e. $v\in \Hcal_+$), the current $J$ behaves as
\be\label{eq:asympJ}
J\sim \tau^{-3} v\rho_{\pm}(v)\,,\quad \tau\rightarrow \pm\infty\,.
\ee
In this case
\be\label{eq:VJ}
V_J(\pm\infty,\bl)=\int_{\Hcal_\pm} \frac{v\rho_{\pm}(v)}{v\cdot \bl} d\mu(v)\,,
\ee
so $V_J(+\infty,\bl)$ need not vanish! (in contrast to the asymptote of the free field, see \eqref{eq:vanishing}). 

The total field decomposes as
\be\label{QED:decomp}
A=A^{\rm R}+A^{\rm in}=A^{\rm A}+A^{\rm out}\,.
\ee
Clearly, $A^{\rm in/out}$ have to solve the homogeneous equation \eqref{wave}, so they are free fields and (assuming that incoming and outgoing fields satisfy the fall-off conditions \eqref{Afaloff1} and \eqref{Afaloff2}), we have the following identities for the asymptotes:
\begin{align*}
V(s,\bl)&=V_J(s,\bl)+V^{\rm in}(s,\bl)=V_J(+\infty,\bl)+V^{\rm out}(s,\bl)\\
V'(s,\bl)&=V_J(-\infty,\bl)+{V^{\rm in}}'(s,\bl)=V_J(s,\bl)+{V^{\rm out}}'(s,\bl)
\end{align*}
Hence
\[
V(+\infty,\bl)=V_J(+\infty,\bl)\,,\qquad V'(-\infty,\bl)=V_J(-\infty,\bl)\,,
\]
which means that $V(+\infty,\bl)$ and $V'(-\infty,\bl)$ come entirely from matter and contribute to the \emph{hard} part of the charge. We also have
\begin{align*}
V(-\infty,\bl)&=V_J(+\infty,\bl)+V^{\rm out}(-\infty,\bl)\,,\\
V'(+\infty,\bl)&=V_J(-\infty,\bl)+{V^{\rm in}}'(+\infty,\bl)\,,
\end{align*}
so both soft and hard components contribute to the matching property that reads:
\be\label{eq:matchingcond}
V'(+\infty,\bl)=V(-\infty,\bl)\,.
\ee

\subsection{Changing the gauge}\label{Sect:gaugetransformation}
In \cite[Chapter 7]{Her16}, one considers a change of gauge:
\[
\hat{A}=A+d\Lambda\,,
\]
where $A$ is a Lorentz potential, but $\hat{A}$ { not necessarily}. The asymptotic field corresponding to $\hat{A}$ is defined by:
\[
\hat{V}_b(s,\bl)=\lim_{R\rightarrow \infty} R\hat{A}_b(R\bl+s\bt/\bt\cdot \bl)\,.
\]
For this limit (and also the limit of $A$) to exist, $\Lambda$ has to be of the form (formula (36) of \cite{Her16}, also confirmed by \cite{CL15}):
\be\label{eq:LambdaExpansion}
\Lambda(R\bl+st/t\cdot \bl)=\varepsilon^+(\bl)+\frac{\beta_{\bt}(s,\bl)}{R}+o(R^{-1})\,,
\ee
which in $u,r,\hat{x}$ variables amounts to: 
\[
\Lambda(x)=\lambda(\hat{x})+O(r^{-1})\,,
\]
where $\lambda(\hat{x})=\varepsilon^+(1,\hat{x})$.
The expansion at time-like infinity  takes the form (following \cite{CL15}):
\be\label{Eq:timelikeinfLambda}
\Lambda(x)=\lambda_{\Hcal}(\rho,\hat{x})+O(\tau^{-\epsilon})
\ee
The resulting contribution to the potential can be computed in the $R,s,\bl$ variables, using the rule \eqref{Rsl}:
\be\label{dLambda}
\frac{\partial}{\partial \bl^b}\Lambda(s\bt/\bt\cdot \bl+R\bl)=\frac{1}{R}\left(V_b^{\varepsilon^+}(\bl)+\bl_b\left(\dot{\beta}_\bt(s,\bl)-\bt\cdot V^{\varepsilon^+}/\bt\cdot \bl\right)\right)
\ee
where $V^{\varepsilon^+}$ is a vector-valued function such that
\[
L_{ab} \varepsilon^+(\bl)=\bl_a V^{\varepsilon^+}_b(\bl)-\bl_b V^{\varepsilon^+}_a(\bl)\,,
\]
and it has properties (see appendix C to \cite{Her16}):
\begin{equation}\label{Vepsilonproperties}
V^{\varepsilon^+}(\lambda \bl)=\lambda^{-1} V^{\varepsilon^+}(\bl)\,,\quad \bl\cdot V^{\varepsilon^+}(\bl)=0\,,\quad L_{[ab}V^{\varepsilon^+}_{c]}(\bl)=0\,.
\end{equation}
Crucially:
\be\label{epsilonV}
\varepsilon^+(\bl)=\frac{1}{4\pi} \int\frac{\bl\cdot V^{\varepsilon^+}(\bl')}{\bl\cdot \bl'}d^2 \bl'\,.
\ee
We also have:
\be\label{epsilonV2}
\int \frac{\varepsilon^+(\bl)}{(\bt\cdot \bl)^2}d^2\bl=\int \frac{\bt\cdot V^{\varepsilon^+}(\bl)}{\bt\cdot \bl}d^2\bl\,.
\ee
\subsection{Green's function}
Following \cite{CL15,CE17}, we consider $\Lambda$ such that $\lambda_\Hcal$ satisfies Laplace equation on the hyperboloid. It is then given in terms of the corner data as
\[
\lambda_{\Hcal}(y)=\int G(y;\hat{x}')\lambda(\hat{x}') d^2\hat{x}'
\] 
where $y$ is the variable at the hyperboloid, $y=(\rho,\hat{x})$ and $G$ is the Green function discussed in \cite{Cam15} with the property that
\[
\lim_{\rho \rightarrow \infty}G(y;\hat{x}')=\delta(\hat{x}-\hat{x}')\,.
\]

\begin{remark}
Observe that we do not require that $\Lambda$ satisfies the wave equation on the whole of $M$. Indeed, this would be incompatible with the observations in \cite{Her16}, reported below in Section \ref{Sect:othergauges}.
\end{remark}

It is also shown in \cite{CL15,CE17} that 
\[
G(y;\hat{x}')=(4\pi)^{-1} (\sqrt{1+\rho^2}-\rho \hat{x}\cdot\hat{x}')^{-2},
\]
while in Herdegen's notation:
\[
\tilde{G}(v;\bl')=(4\pi)^{-1}(\bl'\cdot v)^{-2}\,.
\]
Here $\tilde{G}$ is obtained from $G$, after we set $y=(\rho,\hat{x})$, $\bl'=(1,\hat{x}')$ and $v=(\sqrt{1+\rho^2},\rho \hat{x})$ (i.e. Herdegen's $v$ is $Y$ from \cite{CL15,CE17}, compare with Section \ref{geometricpreliminaries}).
Thus, both references use the same Green's function. Let $\Lambda_{\Hcal}(v)=\lambda_{\Hcal}(y)$.
This allows us to write:
\be\label{eq:LambdaH}
\Lambda_{\Hcal}(v)=\int\tilde{G}(v,\bl)\varepsilon^+(\bl)d^2\bl=\frac{1}{4\pi} \int \frac{\varepsilon^+(\bl)}{(v\cdot \bl)^2} d^2\bl\,.
\ee
Using formula \eqref{epsilonV2}, we obtain
\[
\Lambda_{\Hcal}(v)=\frac{1}{4\pi} \int \frac{v\cdot V^{\varepsilon^+}(\bl')}{v\cdot \bl'}d^2\bl'\,.
\]
As a consistency check, consider the limit 
\[
\lim\limits_{\rho\rightarrow \infty } \int \frac{v\cdot V^{\varepsilon^+}(\bl)}{v\cdot \bl}d^2\bl = \int \frac{l\cdot V^{\varepsilon^+}(\bl')}{l\cdot \bl'}d^2\bl'\,,
\]
where $\bl=(1,\hat{x})$. Using \eqref{epsilonV}, we obtain
\[
\lim\limits_{\rho\rightarrow \infty } \frac{1}{4\pi}  \int \frac{v\cdot V^{\varepsilon^+}(\bl)}{v\cdot \bl}d^2\bl = \varepsilon^+(\bl)\,,
\]
as expected.
\subsection{Lorenz vs. other gauges}\label{Sect:othergauges}
If both $A$ and $\hat{A}$ are in the Lorenz gauge, then (following \cite{Her16}):
\[
\hat{V}(s,\bl)=V(s,\bl)+\bl\alpha(s,\bl)\,.
\]
Assuming $\hat{V}(+\infty,\bl)=V(+\infty,\bl)=0$, it also follows that $\alpha(+\infty,\bl)=0$. In \cite[Section 3.2]{Her16}, it is shown that this implies that
\[
\Lambda(x)=- \frac{1}{2\pi} \int \alpha(x\cdot \bl',\bl')d^2\bl'+\gamma^+\,,
\]
where $\gamma^+$ is a constant\footnote{In \cite{Her16}, $\gamma^\pm$ is denoted $\epsilon_\pm$. We adopt this notation to avoid confusion with $\varepsilon^\pm$, the asymptote of the gauge generator $\Lambda$.}. The null asymptotics are:
\[
\varepsilon^\pm =\lim_{R\rightarrow \infty} \Lambda(st\pm R\bl)=\gamma^\pm\,,
\] 
where 
\[
\gamma^-=\gamma^+- \frac{1}{2\pi} \int \alpha(-\infty,\bl')d^2\bl'\,,
\]
so the matching requirement $\varepsilon^+(\bl)=\varepsilon^-(\bl)$ (see Remark \ref{rem:matching:req}) cannot be met! { This means that if we want the potential $\hat{A}$ to be in the Lorenz gauge and to have non-trivial asymptotics at null infinity, the null asymptotics $\varepsilon^\pm$ of the gauge parameter $\Lambda$ violate the matching requirement. This is a potential issue, since the derivation of the asymptotic charges as presented e.g. in \cite{Strominger,CE17}, assumes all these three properties to hold: Lorenz gauge, non-trivial null asymptotics and the matching requirement. Since the Lorenz gauge is actually only needed asymptotically (so that $\lambda_{\mathcal{H}}$ would satisfy the Laplace equation on the hyperboloid $\mathcal{H}$), we will not impose it in the bulk, so that the no-go result of Herdegen can be circumvented.}

Let us now discuss what happens in non-Lorenz gauges.
Note that in equation \eqref{dLambda}, the contribution from the Lorenz gauge enters the term proportional to $\bl_b$. More generally, the whole term proportional to $\bl_b$ can be absorbed into a residual Lorenz gauge transformation of $V$ and redefinition of $\varepsilon^+(\bl)$. The non-trivial change of the asymptotics is therefore described fully by $V^{\varepsilon^+}(\bl)$ and, following \cite{Her16}, we interpret the resulting transformation (identified as the large gauge transformation of \cite{Strominger}) as
\[
\hat{V}(s,\bl)=V(s,\bl)+V^{\varepsilon^+}(\bl)\,.
\]
The gauge parameter $\Lambda$ used to construct $V^{\varepsilon^+}(\bl)$ does not satisfy the wave equation in the bulk (since $\hat{A}=A+d\Lambda$ cannot be in the Lorenz gauge), but we require that it satisfies it at time-like infinity. We assume $\Lambda$ to be of the form:
\[
\Lambda(x)=\lambda_{\Hcal}(\rho,\hat{x})+f(\tau)\,,
\]
with $f$ vanishing for $\tau\rightarrow \infty$.

\subsection{Calculation of the charge}
We now want to show how a correct specification of boundary fall-off conditions on the fields, and an explicit choice of coordinates around a lightlike boundary $\partial M$ (with corners!) allows us to reproduce known results on asymptotic symmetries in electrodynamics.

The leading idea behind our analysis is that the total asymptotic charge (in the literature \cite{HMPS14,Strominger} it is derived from \emph{large gauge symmetries}\footnote{The notion of large gauge transformation is somewhat ambiguous, as different authors use the same terminology to denote different concepts. Here, it is intended as gauge transformations that do not vanish at infinity.}) is identified as the boundary action $S^\partial$ in the BFV formalism, and the ``charge conservation'' is the consequence of the fact that $S^\partial$ vanishes on-shell. Although there might be deviations from this paradigm, as discussed in Remark \ref{rem:holography}, for the cases at hand we show that the BFV boundary action is the correct functional to consider.

Let us consider a region $\Wcal_R$, whose boundary consists of 4 pieces: $\partial\Wcal_R\doteq\scri_r^+\cup \scri_r^-\cup\Hcal_\tau^+\cup\Hcal_\tau^-$ (cf. Section \ref{geometricpreliminaries}, Figure \ref{diagram}).
Formula \eqref{EMFullBoundaryaction} takes the form
\[
S^{\partial}_{\Wcal_R} = \int_{\partial \Wcal_R} c  \left(-\frac{1}{4\pi}d_A\star F_A  + J\right)
\]
and we immediately observe that, by virtue of Equation \eqref{EOM},
\begin{equation}\label{e:onshellvanishing}
    S^\partial_{\Wcal_R}\approx 0
\end{equation}
for all $R$, where $\approx$ means \emph{on-shell}, i.e. imposing the equations of motion. 

\begin{remark}
Note that, in general, the boundary action does not vanish only by imposing the classical equations of motion, especially if it depends on ghosts and antifields of higher order (see \eqref{dualboundaryaction} below). What we really mean with $\approx$ is taking the degree-zero cohomology of the Koszul--Tate part (the lowest antifield number\footnote{See Section~\ref{sec:fieldsandfunctinals} for the definition.}) of the BV differential. In practice, this is achieved by setting to zero antifields and ghosts for ghosts and quotienting out the EOMs. This demonstrates that the BFV data encodes a large variety  of structural information, which needs to be extracted in the appropriate way.
\end{remark}

We are interested in the limit
\[
S^{\partial}=\lim_{R\rightarrow\infty} S^{\partial}_{\Wcal_R}\,.
\]
Clearly, also $S^\partial\approx 0$ and $S^\partial$ naturally splits into two terms (corresponding to null and time-like asymptotics):
\begin{align}
S^{\partial}&=-\frac{1}{4\pi} \lim_{R\rightarrow \infty}\int_{\partial\Wcal_R} c d_A\star F_A+
\lim_{R\rightarrow \infty}\int_{\partial \Wcal_R} c J\,, \label{EMFullBoundaryactionMoreExplicit}\\
&=\lim_{r\to \infty} S_{\scri_r^+\cup\scri_r^-}^{\partial,\textrm{soft}}+ \lim_{\tau\rightarrow \infty} S^{\partial, \textrm{hard}}_{\Hcal_\tau^+\cup\Hcal_\tau^-}\,, \label{EMFullBoundaryactionExplicit}
\end{align}
where 
\[
S_{\scri_r^+\cup\scri_r^-}^{\partial,\textrm{soft}}=-\frac{1}{4\pi} \int_{\scri_r^+\cup\scri_r^-} c  d_A\star F_A\,,
\]
and
\[
 S^{\partial, \textrm{hard}}_{\Hcal_\tau^+\cup\Hcal_\tau^-}=\int_{\Hcal_\tau^+\cup\Hcal_\tau^-} c J \,.
\]
As noted before (Remark~\ref{rem:cornerterms}), the soft term can also be written as
\be\label{Herdegensway}
S_{\scri_r^+\cup\scri_r^-}^{\partial,\textrm{soft}}=-\frac{1}{4\pi} \int_{\scri_r^+\cup\scri_r^-} d_Ac  \star F_A\,.
\ee

\subsubsection{Soft charge in $(R,s,\bl)$ variables}
We start with the first term in formula \eqref{EMFullBoundaryactionMoreExplicit}. Our assumptions on $A$ imply that the limit is well-defined and we can re-write this term as
\begin{multline*}
    S^{\partial,\textrm{soft}}_{\scri^+\cup\scri^-}
        =\lim_{R\to \infty} S_{\scri_R^+\cup\scri_R^-}^{\partial,\textrm{soft}}
        =-\frac{1}{4\pi}\int_{\scri^+\cup\scri^-}\lim_{R\rightarrow \infty}(R^2d_Ac\wedge \star F_A(x))\\
    =-\frac{1}{4\pi}\int_{\scri^+}\lim_{R\rightarrow \infty}(R^2d_Ac\wedge \star F_A(x))
        -\frac{1}{4\pi}\int_{\scri^-}\lim_{R\rightarrow \infty}(R^2d_Ac\wedge \star F_A(x'))\,,
\end{multline*}
where $x=R\bl + s\frac\bt{\bt\cdot \bl}$ and $x'=-R\bl + s\frac\bt{\bt\cdot \bl}$.
Let's focus on the first terms and evaluate the ghost at the gauge parameter $\Lambda$, as discussed in Remark \ref{Rem:Classicaldegreezerocharges}. We use the fact that
\be\label{eq:limitF}
    \lim_{R\rightarrow \infty}R F_{ab}\left(s\frac\bt{\bt\cdot \bl}+R\bl\right)
    \approx
    \bl_a\dot{V}_b(s,\bl)-\bl_b\dot{V}_a(s,\bl)\,,
\ee
to find the limit of $F_A$ and the expansion \eqref{dLambda} to find the limit of $d_A\Lambda$.

Treating $V^{\varepsilon^+}, \bl, \dot{V}$ as one forms with index lowered using the metric (in the sense that e.g. $\bl^\flat=g(\bl,\cdot)$ is simply written as $\bl$), we obtain:
\begin{multline}
    -4\pi S^{\partial,\text{soft}}_{\scri^+}[\Lambda]=\intl{\scri^+} \lim_{R\to\infty}(R^2 d_A \Lambda \wedge \star F_A) 
    \approx \intl{\scri^+}V^{\varepsilon^+}(\bl)\wedge \star(\bl\wedge \dot{V}(s,\bl)) \\
    = \frac12\intl{\scri^+} V^{\varepsilon^+}_d \epsilon_{abgf}\eta^{gm}\eta^{fn} \bl_m \dot{V}_n dx^a\wedge dx^b\wedge dx^d 
    = \frac12\intl{\scri^+}\mathrm{dVol}_{\scri^+} \epsilon^{rabd}\epsilon_{abgf}\eta^{gm}\eta^{fn} \bl \dot{V}_n\\
    =  \int\limits_{-\infty}^{+\infty}ds\intl{S^2}d^2\bl \left(\delta^{r}_g \delta^d_f -  \delta^{r}_f \delta^d_g\right)\eta^{gm}\eta^{fn} \bl_m \dot{V}_n 
    =  \int\limits_{-\infty}^{+\infty}ds\intl{S^2}d^2\bl \left(\bl^r V^{\varepsilon^+}\cdot \dot{V} - \bl\cdot V^{\varepsilon^+} \dot{V}^r\right)\,,
\end{multline}
where we used that $\bl^r=1$ (see Section \ref{Sect:coordinates}), $dx^a\wedge dx^b\wedge dx^d = \epsilon^{abd}|\mathrm{det}(h)|^{-\frac12}\mathrm{dVol}_{I^+}$ and $h$ is the induced metric on $\scri^+$, the determinant of which is $1$. Observe that $\epsilon^{rabd} \epsilon_{abfg}=2(\delta^{r}_g \delta^d_f -  \delta^{r}_f \delta^d_g)$. Since $\bl\cdot V^{\varepsilon^+}=0$ (see Eq. \eqref{Vepsilonproperties}), and neither $V^{\epsilon^+}$ nor $\bl^r$ depend on $s$, we obtain finally:

\[
    S^{\partial,\text{soft}}_{\scri^+}[\Lambda] 
    = -\frac{1}{4\pi}\intl{\scri^+} \lim_{R\to\infty}(R^2 d_A\Lambda \star F_A)
    \approx
    +\frac{1}{4\pi} \intl{S^2} d^2\bl V^{\varepsilon^+}(\bl)V^{\rm out}(-\infty,\bl)\equiv  Q^{{\rm soft}+}_{\varepsilon^+}\,,
\]
and similarly, the contribution from $\scri^-$ gives:
\[
    S^{\partial,\text{soft}}_{\scri^-}[\Lambda] 
    \approx -\frac{1}{4\pi}\intl{S^2} d^2\bl V^{\varepsilon^-}(\bl){V'}^{\rm in}(+\infty,\bl)\equiv -Q^{{\rm soft}-}_{\varepsilon^-}\,.
\]
So, assuming the matching requirement
\be\label{mc}
\varepsilon^+(\bl)=\varepsilon^-(\bl)\equiv \varepsilon(\bl)\,,
\ee
 the ``soft'' contribution to the boundary action takes the form:
\[
S^{\partial,\textrm{soft}}_{\scri^+\cup\scri^-}[\Lambda]\approx Q^{{\rm soft}+}_{\varepsilon}- Q^{{\rm soft}-}_{\varepsilon}\,.
\]

\begin{remark}
Here we have defined \emph{charges} from the evaluation of the boundary action on different boundary components following the convention that along the past infinity boundary, the sign gets reversed: $S_{\scri^\pm}^{\partial,\text{soft}}=: \pm Q^{\text{soft}\pm}_\epsilon$. The on-shell vanishing of the boundary action is then directly linked to on-shell conservation of charges.
\end{remark}

\begin{remark}\label{rem:matching:req}
Note that the matching \textit{property} \eqref{eq:matchingcond} of the asymptotic potential \cite{Her95} (see (2.26) and the following discussion) is a consequence of equations of motion and the fall-off condition. In contrast to that, the matching \textit{requirement} \eqref{mc} is an extra condition imposed on gauge parameters. It is not a priori clear if this condition can be fulfilled. In fact, it was shown in \cite{Her16} that for $A$ in the Lorentz gauge and $\Lambda$ satisfying the wave equation (both with appropriate fall-off condition), this requirement \textit{cannot be met}.
\end{remark}
\subsubsection{Soft charge in retarded coordinates}
To relate this to the results of \cite{Strominger}, we use the retarded coordinates, as described in Section \ref{Sect:coordinates}. From Equation \eqref{EMFullBoundaryactionExplicit}, we compute the ``soft'' contribution to the boundary action: 
\be\label{EMboundaryaction}
S^{\partial,\textrm{soft}}_{\mathcal{W}_R}=-\frac{1}{4\pi}\int_{\Wcal_R}d_Ac\wedge d_A\star F_A = -\frac{1}{4\pi}\intl{M}\mathrm{dVol}_{\Wcal_R}\ \mathrm{div}(\mathbb{X})
\ee
where $\mathbb{X}^\mu= \frac{c}{\sqrt{g}} \partial_{\tau}\left(\sqrt{g}F^{\mu\tau}\right)$. 

More directly, we get (recall that for an abelian group $d_A\star F_A = d\star F_A$)
\begin{equation}\label{EMfiniteboundaryboundaryation}
S^{\partial,\textrm{soft}}_{\partial\Wcal_R}= -\frac{1}{4\pi}\intl{\Wcal_R}d(c d\star F_A)=-\frac{1}{4\pi}\intl{\partial\Wcal_R} \mathrm{dVol}_\sigma \left[\frac{c}{\sqrt{g}} \partial_\lambda(\sqrt{g}g^{\sigma\mu}g^{\lambda\nu}F_{\mu\nu})\right].
\end{equation}

Observe that the integral over $\partial \Wcal_R$ splits into a null boundary part and a hyperboloid part; however, because of the fall-off conditions and the continuity of the field $A$, in the limit for $\tau\to\infty$ the Hyperboloid contribution vanishes (cf. with Equation \eqref{Eq:hyperboloidfall-off}).

Assume $\Lambda\in\Omega^0(M)$ is a gauge parameter as the one introduced in Section \ref{Sect:gaugetransformation}, i.e. $\Lambda(x) = \Lambda\vert_{\scri}(\hat{x}) + O(r^{-1})$. Note that, when restricted to $\scri\simeq \mathbb{R}\times S^2$, $\Lambda$ is constant along the $\mathbb{R}$ direction.\footnote{Recall that $\hat{x}$ is a coordinate parametrising $S^2$.}

Using retarded/advanced light-cone coordinates and complex coordinates on the 2 dimensional sphere 
we get
$$\mathrm{dVol}= 2r^2 drdu_\pm (1 + z\bar{z})^{-2}dzd\bar{z},$$ 
from which, recalling the explicit expression of formulas \eqref{retardedmink} and \eqref{inverseretardedmink} for Minkowski metric, we obtain 
\begin{multline}
	-4\pi S^{\partial,\textrm{soft}}_{\scri^+\cup\scri^-} = -4\pi \lim_{R\to \infty} S_{\scri_R^+\cup\scri_R^-}^{\partial,\textrm{soft}}
	   = \lim_{R\to \infty} \intl{\scri_R^+\cup\scri_R^-} 
		2\gamma_{z\bar{z}}  dzd\bar{z} du \ 
		 r^2 c(z,\bar{z})\Big[ \partial_u F_{u_\pm r} + \\
		+ \frac{1}{\sqrt{g}}\left[\partial_{z}
		\left(\sqrt{g}g^{rr}g^{z\bar{z}}F_{r\bar{z}}\right)
		+ \partial_{z}\left(\sqrt{g}g^{ru_\pm}g^{z\bar{z}}
		F_{u_\pm\bar{z}}\right) + c.c.\right]\Big]\\
	=\lim_{R\to \infty}  \intl{\scri_R^+\cup\scri_R^-}
	    2\gamma_{z\bar{z}} du_\pm dzd\bar{z}\   
	    c(z,\bar{z}) \left[r^2\partial_{u_\pm} F_{u_\pm r} 
		+ \left[\partial_{z}(\gamma^{z\bar{z}}F_{rz} 
		\mp \gamma^{z\bar{z}}F_{u_\pm\bar{z}})+c.c.\right]\right]\\
	=\intl{\scri^+\cup\scri^-} 2\gamma_{z\bar{z}} du_\pm dzd\bar{z}\   
	    c(z,\bar{z})\left[ \partial_{u_\pm} F_{u_\pm r}^{(2)} 
		+ \left[\partial_{z}(\gamma^{z\bar{z}}F_{rz}^{(0)} 
		\mp \gamma^{z\bar{z}}F_{u_\pm \bar{z}}^{(0)})+c.c.\right]\right]\\
	\approx -\intl{\scri^+_-}2\gamma_{z\bar{z}} dzd\bar{z}c(z,\bar{z}) F^{(2)}_{u^+r} 
	    + \intl{\scri^-_+}2\gamma_{z\bar{z}} dzd\bar{z}c(z,\bar{z}) F^{(2)}_{u^-r} \\
        \mp \intl{\scri^\pm}du_\pm 2\gamma_{z\bar{z}} dzd\bar{z}c(z,\bar{z}) 
		\left[\partial_{z}(\gamma^{z\bar{z}}F_{u_\pm\bar{z}}^{(0)})+c.c.\right]\,,
\end{multline}
where we used that $F_{rz}^{(0)}=0$ due to the assumed fall-off conditions (Equation \eqref{Afall-off}), as well as the vanishing property $F^{(2)}_{u^+r}(+\infty,z,\bar{z}) = F^{(2)}_{u^-r}(-\infty,z,\bar{z})=0$ (cf. Equation \eqref{eq:vanishing}). Observe that the combinations $g^{ru_\pm}g^{u_\pm r}$, $g^{rr}g^{z\bar{z}}$ and $g^{r u_\pm}g^{z\bar{z}}$ have the same sign regardless of the choice of signature for Minkowski metric. 

The matching property (see Equation \ref{eq:matchingcond}) $F^{(2)}_{u^+r}(-\infty,z,\bar{z}) = F^{(2)}_{u^-r}(+\infty,z,\bar{z})$ implies that
\begin{align}\notag
    0&\approx -4\pi S^{\partial, \textrm{soft}}_{\scri^+\cup \scri^-}[\Lambda]\\\notag
        &\approx -\intl{\scri^+}du_+\gamma_{z\bar{z}}dzd\bar{z}
            \Lambda\partial_{z}(\gamma^{z\bar{z}}F_{u_+\bar{z}}^{(0)}) 
        +\intl{\scri^-}du_-\gamma_{z\bar{z}}dzd\bar{z}
            \Lambda\partial_{z}(\gamma^{z\bar{z}}F_{u_-\bar{z}}^{(0)}) + c.c.\\\label{minusSequation}
        &\equiv -4\pi \left[Q_\epsilon^{+\textrm{soft}} - Q_\epsilon^{-\textrm{soft}}\right]\,,
\end{align}
which proves the conservation of the charge and 
is tantamount to the calculations presented in \cite{Strominger}.

\subsubsection{Soft charge from the BV-BFV perspective}
To summarise, in the previous two subsections we have shown that, in the absence of matter, the vanishing of the boundary action on shell (Equation \eqref{e:onshellvanishing}) implies the conservation of the soft charge:
\be
    Q_\epsilon^{+\textrm{soft}} \approx Q_\epsilon^{-\textrm{soft}}.
\ee
This suggest that the result is then independent of the coordinates chosen and the parametrisation of the asymptotic fields. We will clarify this statement in Section \ref{EMcorners}, where the asymptotic symplectic structures in the two parametrisations will be compared.

\subsubsection{Hard charge in $(R,s,\bl)$ coordinates}
For the computation of the ``hard'' charge we assume that the current $J$ has asymptotic behaviour determined by \eqref{eq:asympJ}. The corresponding contribution to the boundary action is given  by:
\[
 S^{\partial, \textrm{hard}}_{\Hcal^+\cup\Hcal^-}=
\lim_{R\rightarrow \infty}\int_{\partial \Wcal_R} c J.
\]
Evaluating the ghost at $\Lambda$, we obtain
\[
 S^{\partial, \textrm{hard}}_{\Hcal^+\cup\Hcal^-}[\Lambda]=\int_{\Hcal^+} (\lim\limits_{\tau\rightarrow\infty}\Lambda)(v) \rho_+(v) dv - \int_{\Hcal^-} (\lim\limits_{\tau\rightarrow\infty}\Lambda)(v) \rho_-(v) dv,
\]
where the relative sign comes from the parametrisation of $\Hcal^-$. Using the asymptotic expansion of $\Lambda$ at time-like infinity (Equation \eqref{Eq:timelikeinfLambda}), this becomes:
\[
 S^{\partial, \textrm{hard}}_{\Hcal^+\cup\Hcal^-}[\Lambda]=
\int_{\Hcal^+}\Lambda_{\Hcal^+}(v) \rho_{+}(v) dv - \int_{\Hcal^-}\Lambda_{\Hcal^-}(v) \rho_-(v) dv\,,
\]
so applying \eqref{eq:LambdaH}, we obtain:
\begin{multline*}
 S^{\partial, \textrm{hard}}_{\Hcal^+\cup\Hcal^-}[\Lambda]=
\frac{1}{4\pi} \int \frac{\varepsilon^+(\bl)}{(v\cdot \bl)^2}\rho_+(v) d^2\bl dv- \frac{1}{4\pi} \int \frac{\varepsilon^-(\bl)}{(v\cdot \bl)^2}\rho_-(v) d^2\bl dv\\=\frac{1}{4\pi} \int \frac{v\cdot V^{\varepsilon^+}(\bl)}{v\cdot \bl}\rho_+(v)d^2\bl dv-\frac{1}{4\pi} \int \frac{v\cdot V^{\varepsilon^-}(\bl)}{v\cdot \bl}\rho_-(v)d^2\bl dv\,,
\end{multline*}
where in the second step we used the identity \eqref{epsilonV2}. Assuming again the matching condition \eqref{mc}, we obtain:
\[
 S^{\partial, \textrm{hard}}_{\Hcal^+\cup\Hcal^-}[\Lambda]=Q^{{\rm hard}+}_{\varepsilon}-Q^{{\rm hard}-}_{\varepsilon}\,,
\]
where
\[
Q^{{\rm hard}+}_{\varepsilon}=\frac{1}{4\pi} \int \frac{v\cdot V^{\varepsilon}(\bl)}{v\cdot \bl}  \rho_+(v)d^2\bl dv=\frac{1}{4\pi} \int V^{\varepsilon}(\bl)\cdot V_J(+\infty,\bl) d^2\bl\,,
\]
where we inserted the expression for $V_J$ given by equation \eqref{eq:VJ}. Similarly:
\[
Q^{{\rm hard}-}_{\varepsilon}=\frac{1}{4\pi} \int  V^{\varepsilon}(\bl)\cdot V_J(-\infty,\bl) d^2\bl\,,
\]
and the total contribution from the hard charge is:
\[
Q^{{\rm hard}}_{\varepsilon}=Q^{{\rm hard}+}_{\varepsilon}-Q^{{\rm hard}-}_{\varepsilon}.
\]

\subsubsection{Hard charge in retarded coordinates}
We use a coordinate system adapted to the $\tau$-hyperboloid part of the boundary of $\Wcal_R$ (see Section \ref{sec:hyperbolic}), and then take a limit for $\tau \to \infty$. Observe that the fall-off conditions for the current $J$ imply that the terms on $\scri_R^\pm$ vanish in the limit $R\to\infty$, so we can discard them from the outset. We compute
\begin{align*}
    S^{\partial, \textrm{hard}}_{\Hcal^+\cup\Hcal^-}[\Lambda] & = \lim_{\tau\to\infty}S^{\partial, \textrm{hard}}_{\Hcal_\tau^+\cup\Hcal_\tau^-}[\Lambda] 
        =\lim_{\tau\to\infty}\intl{\Hcal_\tau^+\cup\Hcal_\tau^-}\mathrm{dVol}_{\Hcal_\tau^\pm} \Lambda\vert_{\Hcal^\pm}J\\
    & =\lim_{\tau\to\infty}\intl{\Hcal_\tau^+\cup\Hcal_\tau^-} \mathrm{dVol}_{\Hcal^\pm}\tau^{3}J \int d^2\hat{x}' G(y;\hat{x}')\varepsilon(\hat{x}')    \\
    & =\intl{\Hcal^+\cup\Hcal^-} \mathrm{dVol}_{\Hcal^\pm}\int d^2\hat{x}'G(y;\hat{x}')J^{(3)}\varepsilon(\hat{x}')    \\
    &=\frac{1}{4\pi} \int d^2\hat{x}' \intl{\Hcal^+\cup\Hcal^-} \mathrm{dVol}_{\Hcal^\pm}\frac{Y \varepsilon(\hat{x}') }{(Y\cdot \hat{x}')^2}\rho_\pm(Y) \\
    &=\frac{1}{4\pi} \int d^2\hat{x}' \intl{\Hcal^+\cup\Hcal^-} \mathrm{dVol}_{\Hcal^\pm}\frac{Y\cdot V^{\varepsilon}}{Y\cdot \hat{x}'}\rho_{\pm}(Y)\\
    &= Q^{\textrm{hard}+}_\varepsilon - Q^{\textrm{hard}-}_\varepsilon\,,
\end{align*}
where we used Equation \eqref{eq:asympJ} to rewrite $J^{(3)}$ and Equation \eqref{epsilonV2} between lines 4 and 5.
\subsubsection{Total charge from the BV-BFV perspective}
The total charges are given by:
\begin{multline*}
Q^{+}_{\varepsilon}=Q^{{\rm soft}+}_{\varepsilon}+Q^{{\rm hard}+}_{\varepsilon}\approx \frac{1}{4\pi} \int d^2\bl V^{\varepsilon}(\bl)(V_J(+\infty,\bl)+V^{\rm out}(-\infty,\bl))\\=\frac{1}{4\pi} \int d^2\bl V^{\varepsilon}(\bl)V(-\infty,\bl)
\end{multline*}
and
\begin{multline*}
Q^{-}_{\varepsilon}=Q^{{\rm soft}-}_{\varepsilon}+Q^{{\rm hard}-}_{\varepsilon}\approx \frac{1}{4\pi} \int d^2\bl V^{\varepsilon}(\bl)(V_J(-\infty,\bl)+{V'}^{\rm in}(+\infty,\bl))\\=\frac{1}{4\pi} \int d^2\bl V^{\varepsilon}(\bl)V'(+\infty,\bl)\,,
\end{multline*}
so vanishing of the boundary action in the BV-BFV formalism implies the charge conservation
\[
Q^{-}_{\varepsilon}\approx Q^{+}_{\varepsilon}\,.
\]

{
\subsection{Symplectic BFV analysis}\label{EMcorners}
Recall that the BFV structure comes equipped with a symplectic form $\Omega^\partial$ on a graded manifold of boundary fields $\mathcal{F}^\partial$, which is related to the canonical $(-1)$-symplectic BV form $\Omega$ by
\begin{equation}
    \mathcal{L}_Q \Omega = \pi^* \Omega^\partial.
\end{equation}
For Yang--Mills theory (and in particular Electrodynamics) with scalar sources this reads
\begin{equation}
    \Omega^\partial = \intl{\partial M} \frac{1}{4\pi}\delta A \delta[\star F_A] + \delta A^\ddag \delta c + \delta \oph \delta [\star d_A\ph] + c.c.
\end{equation}
and $Q^\partial$ is the Hamiltonian vector field of $S^\partial$ with respect to $\Omega^\partial$. Notice that we treat $[\star F_A]$ restricted to the boundary as an independent field. {The first term is equal on-shell to the standard symplectic form used in asymptotic quantization of the electromagnetic field \cite{Ash80}. In \cite{Her96}, this symplectic form is refereed to as \textit{radiated symplectic form}, so we will use the notation $\Omega_{\rad}$ for it and distinguish it from $\Omega_{\her}$, which is the symplectic form used by Herdegen (\emph{ibid.}).

Recall that, in the notation of \cite{Her95,Her96,Her16}, the asymptotic variables on $\scri^+$ are $V$ and $\dot{V}$; the symplectic form in terms of these variables is given by
\begin{equation}\label{AQform}
\Omega_{\her}=\frac{1}{4\pi}\int\limits_{-\infty}^{+\infty}ds\intl{S^2}d^2\bl\,  \delta V\cdot \delta \dot{V}\,.
\end{equation}
To compare this with $\Omega_{\rad}=\intl{\partial M} \frac{1}{4\pi}\delta A \delta[\star F_A]$, we consider the following limit:
\begin{equation}
\lim_{R\rightarrow\infty}\intl{\scri^+_R} \frac{1}{4\pi} \delta A \delta[\star F_A]=\intl{\scri^+}\lim_{R\rightarrow\infty} \frac{R^2}{4\pi} \delta A \delta[\star F_A],
\end{equation}
where we have extracted $R^2$ from the volume form on $\scri^+_R$, i.e. $\mathrm{dVol}_{\scri^\pm_R} = R^2 \mathrm{dVol}_{\scri^\pm}$.
Using equation \eqref{eq:limitF} and the reasoning presented in the paragraph following that equation, we conclude that
\begin{align*}
\lim_{R\rightarrow\infty}\intl{\scri^+_R} \frac{1}{4\pi} \delta A \delta[\star F_A]
    &=\intl{\scri^+} \lim_{R\rightarrow\infty} 
        \frac{R^2}{4\pi} \delta A \delta[\star F_A] \approx\intl{\scri^+}\delta V(\bl)\wedge \delta(\star(\bl\wedge \dot{V}(s,\bl))) \\
    &=  \int\limits_{-\infty}^{+\infty}ds\intl{S^2}d^2\bl \left(\bl^r \delta V\cdot \delta \dot{V} - \bl\cdot \delta V \delta \dot{V}^r\right)\\
    &=\int\limits_{-\infty}^{+\infty}ds\intl{S^2}d^2\bl\,  \delta V\cdot \delta \dot{V}-\int\limits_{-\infty}^{+\infty}ds\intl{S^2}d^2\bl  (\delta V\cdot\bl) \delta \dot{V}^r,
\end{align*}
since $\bl^r=1$, and both $A$ and $\star F_A$ have $1/R$ leading terms at $\scri^\pm$. Now we use the fact that $V\cdot \bl$ is, on-shell, the total charge of the system, { which vanishes in the chargeless case (see e.g. formula (2.3) of \cite{Her96}). Then, variations on the vector space of chargeless asymptotes must also be chargeless, i.e. $\bl\cdot \delta V =0$, making this expression the same as \eqref{AQform}:
\[
\lim_{R\rightarrow\infty}\intl{\scri^+_R} \frac{R^2}{4\pi} \delta A \delta[\star F_A] = \Omega_{\her}.
\]
A similar reasoning can be also repeated for $\scri^-$.  Then, the space of (chargeless) asymptotes $V(s,\bl)$ and vector valued functions $\dot{V}(s,\bl)$, with symplectic form given by $\Omega_{\her}$, is identified (on shell) with the space of boundary fields $\{A\vert_{\partial M},[\star F_A]_{\partial M}\}$ with symplectic form $\Omega_{\rad}$.
}
\begin{remark}
To treat the charged case, Herdegen is also using $\Omega_{\her}$ rather than $\Omega_{\rad}$, even though they differ in this case. The proper way to treat it from the geometrical perspective is to work with affine spaces, to fix the reference connection and then consider perturbation that are either charged or chargeless, depending on the choice of field space topology.
\end{remark}

Observe that one can compute the canonical brackets associated to $\Omega_{\her}$ after applying a large gauge transformation (LGT) on (on-shell) bulk field configurations\footnote{Seen as functions of the initial (asymptotic) values $\{V,\dot{V}\}$.} $\hat{V}_i = V_i + V_i^{\varepsilon^+}$. Following \cite{Her16}, one gets
\begin{equation}\label{herdegenfailure}
\{\hat{V}_1,\hat{V}_2\}= \{V_1,V_2\} + \intl{S^2}d^2\bl V_1^{\varepsilon^+}\cdot V_2(-\infty,\bl)  -\  \{1\leftrightarrow 2\}.
\end{equation}}
If $X_\varepsilon=V^{\varepsilon^+}\frac{\delta}{\delta V}$ denotes the LGT, this is recovered as:
\begin{equation}\label{herdegensymplecticfailure}
    \mathcal{L}_{X_{\varepsilon}}\Omega_{\her}=\int\limits_{-\infty}^{+\infty} ds\intl{S^2}d^2\bl \delta V^{\varepsilon^+}\cdot\delta\dot{V} \approx \intl{S^2}d^2\bl \delta V^{\varepsilon^+}\cdot\delta V(-\infty,\bl),
\end{equation}
where we used that, on shell, $\dot{V}(s,\bl)=\partial_sV(s,\bl)$, and that $\partial_s V^{\varepsilon^+}=0$ to identify a total $s$-derivative, while $V(+\infty,\bl)$ vanishes in the sourceless case. Formula \eqref{herdegenfailure} is obtained by \eqref{herdegensymplecticfailure} by simply evaluating on two on-shell configurations. 

The failure of invariance of the canonical structure under Large Gauge Transformations is argued to mean that LGT's are ``not symmetries of the asymptotic structure''. We will show now that from the viewpoint of BV-BFV the failure of invariance of $\Omega^\partial$ is in fact expected and fully consistent with the structure, if one appropriately takes into account the corner data.  
 Indeed, it is a straightforward calculation to check that $\mathcal{L}_{Q^\partial} \Omega^\partial$, as anticipated in Section \ref{Corners}, is a total derivative (off-shell): 
\begin{equation}\label{bdrformfailure}
    \mathcal{L}_{Q^\partial} \Omega^\partial = \intl{\partial M} d\left( \delta c \delta [\star F_A]\right).
\end{equation}
{ To compare \eqref{bdrformfailure} with Equation \eqref{herdegensymplecticfailure}, we use \eqref{eq:limitF} again. For simplicity, we consider the free theory (i.e. $J=0$). Let $\Lambda$ be a gauge parameter implementing the given LGT, so that $A$ transforms to $A+d\Lambda$, as discussed in Section \ref{Sect:othergauges}. Then the $\scri^+$ contribution to the invariance-breaking term takes the form:
 \begin{multline}\label{cornersympHer}
\lim_{R\rightarrow\infty} \intl{\scri^+_R}R^2 d\left( \delta c \delta [\star F_A]\right)[\Lambda]
        \approx  
        -\lim_{R\rightarrow\infty} \intl{\scri_R^+} R^2\, \delta (d\Lambda)\wedge \delta [\star F_A]\\
    =\intl{\scri^+} \lim_{R\to\infty}(R^2 \delta (d\Lambda) \wedge \delta(\star F_A)) 
        \approx 
        \intl{\scri^+}\delta(V^{\varepsilon^+})(\bl)\wedge\delta(\star(\bl\wedge \dot{V}(s,\bl)))\\
    =  -\int\limits_{-\infty}^{+\infty}ds\intl{S^2}d^2\bl\, \delta V^{\varepsilon^+}\cdot \delta\dot{V}
        \approx 
        \intl{S^2}d^2\bl\, \delta V^{\varepsilon^+}(\bl)\cdot\, \delta V(-\infty,\bl)\,,
\end{multline}
where we have commuted $\delta$ and $d$ and we used the fact that in the sourceless case $d\star F_A\approx 0$ in the first line, while the explicit asymptotic expressions $V^{\epsilon^+}$ and $\bl\wedge \dot{V}$ for $d\Lambda$ and $F_A$ respectively were used to obtain the second line. The last line is equivalent to \eqref{herdegensymplecticfailure}. The breaking of the invariance is related to the fact that in general $V(-\infty,\bl)$ does not vanish. For the sourced case, this invariance-breaking is characterized by $\Delta V(\bl)= V(-\infty,\bl)-V(+\infty,\bl)$ (using the notation of \cite{Her16}). }

We can interpret the final expression in Equation \eqref{herdegensymplecticfailure} as a symplectic structure on the space of ``chargeless'' variations $V^{\epsilon^+}$ at the asymptotic corner, i.e. such that $\bl\cdot V^{\epsilon^+}=0$, and on-shell variations of the asymptotic field $V$. The canonical Poisson brackets associated to this symplectic form compute exactly the failure of Equation \eqref{herdegenfailure}.

Alternatively, starting again from \eqref{bdrformfailure} we can compute (off-shell) 
\begin{equation}\label{cornersympStro}
    \lim_{R\rightarrow\infty} \intl{\scri^+_R}R^2 d\left( \delta c \delta [\star F_A]\right)[\Lambda] = \lim_{R\rightarrow\infty} \intl{\partial \scri^+_R} R^2 \delta \Lambda \delta [\star F_A] = \intl{\scri^+_-\cup\scri^+_+} d^2\bl\,\delta\varepsilon(\bl)\, \delta F_{ru}^{(2)}(\bl).
\end{equation}
This means that we can identify two symplectic manifolds on the corner. The first one is the space of vector valued functions $V^{\epsilon^\pm}$ orthogonal to $\bl$ (i.e. chargeless) and asymptotes $V(\mp\infty,\bl)$, with symplectic form given by Equation \eqref{cornersympHer}. The second one is given by the space of functions $\varepsilon(\bl)$ on $S^2$ together with the space of asymptotes $F^{(2)}_{ru}$, with symplectic form given by Equation \eqref{cornersympStro}. The fact that they are both obtained from Equation \eqref{bdrformfailure} means that, on-shell, there is a symplectomorphism between these two spaces, given by Equations \eqref{epsilonV} --- relating $\varepsilon^\pm(\bl)$ to $V^{\varepsilon^\pm}(\bl)$ --- and \eqref{eq:limitF}, which in turn relates $F^{(2)}_{ru_\pm}$ to $V(\pm\infty,\bl)$.

\begin{remark}
The expression on the right hand side of Equation \eqref{cornersympStro} is the infinitesimal version of the corner symplectic structure identified by Donnelly and Freidel in \cite{DF16}. One obtains their exact expression by considering finite gauge transformations $A\to g^{-1} A g + g^{-1} dg$ for $g$ a group-valued function. In their work, gauge invariance is restored by means of an extension of the physical phase space (and gluing is achieved through fusion of symplectic data). Despite the slight difference in the underlying philosophy, the detected phenomenon is the same. The asymptotic BV-BFV analysis contains all the relevant physical information, and packages it in a way suitable for quantisation \cite{CMR2}.
\end{remark}

Let's now come back to the BV-BFV interpretation of Equation \eqref{bdrformfailure}. If we denote the corner manifold by $K$, we can denote the space of corner fields by $\mathcal{F}^{\partial\partial}_{K}= C^\infty[1](K,\mathfrak{g})\oplus \Omega^2(K,\mathfrak{g}^*)$. There is a natural surjective submersion $\pi_\partial$ from the space of boundary fields to $\mathcal{F}^{\partial\partial}_{K}$ given by restriction of fields\footnote{Technically, this is done in two steps: restrictions of fields yields a pre-symplectic manifold, which then needs to be reduced to yield $F^{\partial\partial}_K$. For theories such as Yang--Mills this step is almost trivial.}, and the right hand side of formula \eqref{bdrformfailure} can be interpreted as the canonical one form on $\mathcal{F}^{\partial\partial}_{K}$, denoted by $\Omega^{\partial\partial}$. Then we have
$$
\mathcal{L}_{Q^\partial} \Omega^\partial = \pi^*\Omega^{\partial\partial}.
$$
Equation \eqref{cornersympStro} computes the asymptotic limit of the corner symplectic form $\Omega^{\partial\partial}$. Observe that if we denote the projection to nonvanishing asymptotic corner fields by $\pi_\infty$, the structural BV-BFV relations hold:
\begin{equation}
    \mathcal{L}_{Q^\partial}\Omega^\partial = \pi_\infty^*\Omega^{\partial\partial}_\infty
\end{equation}
where $\Omega^{\partial\partial}_\infty$ is the r.h.s. of Equation \eqref{cornersympStro}. The  BV-BFV axioms require the  map $\pi_\infty$ to be a surjective submersion. This can be shown easily if the space of fields is a vector space, by multiplying the desired asymptotic with an appropriate homogeneous function and a compactly supported function vanishing at 0. In general this would depend on the choice of field space topology (and hence the ``size'' of tangent space) and has to be verified.

Notice that $\mathcal{L}_{Q^\partial}$ is a differential on the space of boundary fields --- the BFV operator. Therefore, Equation \eqref{bdrformfailure} implies that, although the symplectic form fails to be a cocycle for the BFV differential, such failure is controlled by a symplectic form (of degree $1$) associated to a space of corner fields (cf. Section \ref{extgauge}).

In the presence of higher codimension strata we can find a solution of the descent equation, i.e. an inhomogeneous local form valued density $\varpi^\bullet$:
\begin{equation}
    \varpi^\bullet = \delta A^\ddag \delta A + \delta c\delta c^\ddag + \delta A\delta [\star F_A] + \delta A^\dag \delta c + \delta c \delta [\star F_A],
\end{equation}
which satisfies
\begin{equation}
    (\mathcal{L}_Q -d) \varpi^\bullet = 0.
\end{equation}
Equation \eqref{formsfromdensity} tells us that we can recover the symplectic forms at every codimension by restriction of $\varpi^\bullet$ and integration over the appropriate submanifold (stratum).

}

\subsection{Summary}\label{sec:dis}
By applying the classical BV-BFV formalism \cite{CMR1} to electrodynamics, we have shown how the on-shell vanishing of the boundary action leads to the existence of a conserved quantity $Q_{\varepsilon}$, akin to a Noether charge, which in the recent literature (see e.g. \cite{Strominger}) is suggested to be the generator of a large gauge transforamtion. On the other hand, \cite{Her16} it is argued that the transformation in question is not a symmetry, but rather a map into a different sector of the theory. 

We agree with the latter claim of \cite{Her16} according to which, in order to derive the conservation of $Q_{\varepsilon}$, one needs to consider transformations of the gauge potential that lead away from the Lorentz gauge. However, we also show that by appropriately choosing fallof conditions for such transformations, one can reproduce the formulas derived in \cite{CL15,Strominger,CE17}. Indeed, we have shown that $Q_{\varepsilon}$ is computed by the boundary action $S^\partial$, seen as the BFV version of the Noether charge, for gauge transformations with nonvanishing asymptotics. Hence our interpretation is also close to the one of \cite{Strominger}.

To give an answer to our initial question ``in which sense are LGT's symmetries of the theory'' we have performed an analysis of the canonical symplectic structure of Electrodynamics, and showed how its failure under large gauge transformations fits naturally in the BV-BFV language. This suggests an extension of the notion of symmetry for a gauge field theory, which agrees with the philosophy that led to the study of descent equations. In other words, large gauge transformations are \emph{extended symmetries} of the theory, completely encoded in higher codimension structural data of the field theory, and possibly fundamental for a correct, covariant and functorial quantisation of the theory  \cite{CMR2}.

Important implications for the quantum theory follow from the fact that LGTs of \cite{CL15,Strominger,CE17}  are transformations between theories in different gauges. Indeed the results of \cite{DybWe19} suggest that different gauges lead to unitarily in-equivalent theories, i.e. representing different sectors, as also stated in \cite{Her16}. We plan to say more about the behaviour of the quantised theory upon the choice of gauge in our future work.

\section{The Scalar Field Theory}\label{Sect:Scal}
In this section we discuss the case of asymptotic symmetries in the scalar field. Inspired by the work in \cite{CC,CFHS}, we will show how conserved asymptotic charges for the scalar field can be obtained from the BV-BFV approach to a ``dual'' two-form model. In doing so, we extend the standard duality of free models to the sourced scenario. This provides an alternative to the analysis presented in \cite{CFHS}, which instead considered symmetries of the dual model to be given by elements of the cohomology group $H^2(M)$. In Section~\ref{s:LGTnoBV} we show that this type of symmetry (akin to constant shifts of a scalar field) does not admit a BV description.

Throughout, we consider $(M,g)$ to be a closed, $4$-dimensional Lorentzian manifold with boundary, and the \emph{space of classical fields}\footnote{One could as well consider complex-valued scalars.} is $F_{cl}\coloneqq C^\infty(M,\mathbb{R})\ni\phi$. The classical \emph{action functional} is given by
\begin{equation}
	S_{cl}=\intl{M}\star_g d\phi \wedge d\phi,
\end{equation}
where $\star_g$ is the Hodge-dual operator defined by the pseudo-Riemannian metric $g$.

Free scalar field theory is \emph{classically equivalent} to a theory of 2-forms $\B\in\Omega^2(M)$
\begin{equation}
    S_{\text{dual}}=\intl{M} d\B \star_g d\B,
\end{equation}
meaning that, on-shell, we can set $d\B \coloneqq \star_g d\phi$ and --- up to symmetries --- we obtain a diffeomorphism of the spaces of solutions of the Euler Lagrange equations of the two models. In what follows we will drop the subscript $g$ and denote $\star_g\equiv \star$.

\begin{remark}\label{dualityremark}
To be precise, note that the equation of motion $d\star d\phi=0$ is the statement that the three-form $\Hsf_\phi\coloneqq \star d\phi$ is closed. Assuming that $H^3(M)=0$, we can then find $\B\in\Omega^2(M)$ such that $\Hsf_\phi=d\B$, as done above. This is clearly the case on Minkowski space $M=\mathbb{M}^4$, but it is in general not true, and the condition $H^3(M)=0$ needs to be checked.
\end{remark}

{
Notice that the ``duality'' between a free scalar field theory and a free two-form model is incoded in the pair of equations:
\begin{equation}\label{nonsourcedduality}
    \begin{cases}
        d\star d \B =0 \iff d^2 \phi=0 \\
        d\star d \phi =0 \iff d^2 \B=0
    \end{cases}
\end{equation}
toggling between the Bianchi identity and field equations. We will see in Section \ref{Sec:hardscalar} how to extend this to field equations with nontrivial external sources.
}
\begin{remark}
Clearly the definition of $\B$ is not unique, and the theory enjoys a symmetry $\B \to \B+ d\gamma$. Notice that $\gamma$ is also defined up to an exact form, and thus enjoys an additional symmetry $\gamma\to\gamma + d\tau$. This is an example of a reducible symmetry, which can be easily treated in the BV formalism.
\end{remark}

In what follows we will analyse the dual field theory in the BV formalism, and then recover the ``soft'' scalar charges of \cite{CCM}. We assume that $H^3(M)=0$.

\subsection{BV-BFV analysis of the dual model}\label{Sect:BVdualmodel}
The dual model has a built-in symmetry given by rescaling $B$ by a closed form $\beta$. The standard way to proceed here would be considering the ``gauge'' symmetry of $\B$ in terms of exact forms $\B\to \B + d\gamma$, and extend the dual model to the BV setting. 

A different point of view was proposed in \cite{CFHS}, which relies on symmetries $\B\to \B + \beta$, generated by $\beta\in H^2(M)$, i.e. a closed but not exact form on $M$. We will turn to this latter possibility --- and the complications that arise --- in Section \ref{s:LGTnoBV}, after we have analysed the standard case. 

Notice that in $\B\to \B + d\gamma$, the form $\gamma$ also enjoys a symmetry, as we can freely map $\gamma\to \gamma + d\tau$. The BV formalism produces the extended action functional 
\begin{equation}
    S_{\text{dual}}^{BV} = \intl{M} d\B \star d\B + \B^\ddag d\gamma + \gamma^\ddag d\tau,
\end{equation}
and the BV operator  $\iota_Q\Omega_{\text{dual}} = \delta S_{\text{dual}}^{BV}$
\begin{eqnarray}
    Q\B = d\gamma 
        & Q\gamma=d\tau 
            & Q \tau =0\\
    Q\B^\ddag = d\star d\B 
        & Q\gamma^\ddag = d\B^\ddag 
            & Q\tau^\ddag = d\gamma^\ddag 
\end{eqnarray}
that satisfies $[Q,Q]=2Q^2=0$, on the space of fields:
$$
    \mathcal{F}_{\text{dual}}\coloneqq T^*[-1]\left(\Omega^0(M)[2]\times \Omega^1(M)[1]\times \Omega^2(M)\right)\ni (\tau^\ddag, \gamma^\ddag, \B^\ddag, \B, \gamma, \tau).
$$
If $M$ has a non-empty boundary, it is easy to check that the boundary one-form
\begin{equation}
    \alpha^\partial_{\text{dual}}\coloneqq \intl{\partial M} \delta \B \Hsf_{\B} + \B^\ddag \delta \gamma + \gamma^\ddag \delta \tau
\end{equation}
with $\Hsf_{\B}\coloneqq \star d\B\vert_{\partial M}$, and the boundary action
\begin{equation}\label{dualboundaryaction}
    S^\partial_{\text{dual}}\coloneqq\intl{\partial M}   d\gamma \Hsf_{\B} + d\tau \B^\ddag
\end{equation}
satisfy the BV-BFV axioms, namely
\begin{equation}
    \iota_Q\Omega_{\text{dual}} = \delta S_{\text{dual}}^{BV} + \pi^*\alpha^\partial_{\text{dual}}
\end{equation}
and
\begin{equation}
    \frac12\iota_Q\iota_Q\Omega_{\text{dual}} = \pi^* S^\partial_{\text{dual}},
\end{equation}
{with $\pi$ simply the restriction of fields (and normal jets) to the boundary. }For simplicity of notation, we will drop the subscript ``dual'' in what follows.

\subsection{Asymptotic symmetries of the dual model}\label{Sect:Asymduelmodel}
We would like now to revert to the scalar field description, and use the boundary action found so far as a generator for our asymptotic charges. In what follows, we will reconstruct asymptotic symmetries from the BV-BFV formulas obtained so far, after choosing appropriate fall-off condtions. Observe that the on-shell condition $\star d\B = d\phi$ restricts to the boundary  
$$
    \Hsf_{\B}\equiv\star d \B\vert_{\partial M} 
        \approx d\phi\vert_{\partial M}
$$
and if the boundary component has a boundary of its own, for example a sphere at the corner of a lightlike boundary, we get 
\begin{equation}
    S^\partial\approx \intl{\partial M}d\gamma d\phi\vert_{\partial M} = \intl{\bigcup_{i} \partial (\partial M)^i} d\gamma \phi ,
\end{equation}
where the symbol $\approx$ means that we enforced the equations of motion and set antifields to zero, and $(\partial M)^i$ denotes the $i$-th connected component of the boundary (with appropriate orientation).

Evaluating the boundary action on a specific gauge parameter $\Gamma \in\Omega^1(M)$ (see Remark \ref{Rem:Classicaldegreezerocharges}), we can extract the volume form of $S^2$ from the two-form $d\Gamma$, defining a function $\lambda\in C^\infty(S^2)$ such that 
\begin{equation}
    d\Gamma(x) = \lambda(x)dS^2\,,
\end{equation}
so that the corner term becomes
\begin{equation}\label{cornerchargefromBVdual}
    S^\partial[\Gamma]=\intl{\bigcup_{i} \partial (\partial M)^i}  \lambda(x) \phi dS^2.
\end{equation}
\subsection{Asymptotic symmetries of the free scalar field}\label{sec:Sym:ScalarField}
We would like to discuss now how formula \eqref{cornerchargefromBVdual} produces asymptotic charges for scalar fields. From now on, we  restrict our discussion to asymptotically flat Lorentzian manifolds $M$ and, for simplicity, one can consider Minkowski spacetime\footnote{Possible global effects will not be discussed here.}.

In retarded light-cone coordinates we define a boundary at infinity $\mathcal{I}$ by the condition $r=R\to \infty$. We consider scalar fields with the following radial dependence\footnote{We consider this radial expansion in order to match with \cite{CFHS}.}: 
$$
    \phi=\sum_{k=1}^{\infty} \phi^{(k)} r^{-k}\,,
$$
with $\phi^{(k)}$ independent of $r$, so that 
$$
    d\phi = \frac{1}{r} d\phi^{(1)} 
        + \sum_{k=2}\frac{1}{r^k}d\phi^{(k)} 
        - dr \sum_{k=1} \frac{1}{r^{k+1}}\phi^{(k)}.
$$
Observe that the $dr$ part is obviously not present in the restriction $d\phi\vert_{\partial M}$. From the definition of $\Hsf_\phi=\star d\phi$ we get
$$
    \Hsf_\phi = \sum_{k=1}\frac{1}{r^k} \star d\phi^{(k)} - \star dr \sum_{k=1}\frac{1}{r^{k+1}} \phi^{(k)}\,,
$$
so, requiring that $\Hsf_\phi=d\B$, we are lead to
$$
    \B = \sum_{k=1}^\infty \frac{1}{r^k} \B^{(k)},
$$
and we set the fall-off condition for $\gamma$ to be such that $\gamma=\sum_{k=1}^\infty \frac{1}{r^k}\gamma^{(k)}$.

As in the case of Electrodynamics, we can express the above results in terms of $(R,s,\bl)$ variables. Following \cite{Her95}, we define
\be\label{as1}
\lim\limits_{R\rightarrow \infty} R\ph(x+R\bl)=\chi(x\cdot \bl,\bl)\,,
\ee
and identify
\[
\chi(s,\bl)=\phi_+^{(1)}(u_+,\hat{x})\,.
\]
We assume the fall-off conditions
\be\label{faloff1}
|\chi(s,\bl)|<\frac{const.}{s^\epsilon}\,,
\ee
\be\label{faloff2}
|\dot\chi(s,\bl)|<\frac{const.}{s^{1+\epsilon}}\,,
\ee
so that $\chi(+\infty,\bl)=0$ (i.e. $\phi_+^{(1)}(\infty,\hat{x})\equiv 0$), and recall that in the absence of external currents we have
\be\label{eq:free:scalar}
\Box\phi=0\,.
\ee
Under these assumptions, it was shown in \cite{Her95} that
\be\label{recon1}
\phi(x)=-\frac{1}{2\pi} \int \dot\chi(x\cdot \bl,\bl)d^2\bl\,.
\ee
Now consider the past null asymptotics. Take a homogeneous function $\chi'$ satisfying \eqref{faloff1} and \eqref{faloff2} with $\bl$ replaced by past-pointing null directions $-\bl$. We then have
 \be
 \lim\limits_{R\rightarrow \infty} R\ph(x-R\bl)=\chi'(x\cdot \bl,\bl)\,.
 \ee
 and identify $\chi'(s,\bl)=\phi_-^{(1)}(u_-,\hat{x})$. Fall-off conditions analogous to \eqref{faloff1} and \eqref{faloff2} imply that $\chi'(-\infty,\bl)\equiv 0$ (i.e. $\phi^{(1)}_-(-\infty)=0$).
 
 The field can now be also expressed as:
 \be\label{recon2}
 \phi(x)=\frac{1}{2\pi} \int \dot\chi'(x\cdot \bl,\bl)d^2\bl\,.
 \ee
Comparing \eqref{recon1} and \eqref{recon2} we obtain:
\be
\int (\dot\chi(s,\bl)+\dot\chi'(s,\bl))d^2\bl=0\,.
\ee
It was shown in \cite{Her95} that this in fact implies
\be
\dot\chi(s,\bl)+\dot\chi'(s,\bl)=0\,.
\ee
We obtain the existence of the limits $\chi(-\infty,\bl)$ and $\chi'(+\infty,\bl)$ as well as
\be\label{MatchingProp:scalar}
\chi(s,\bl)+\chi'(s,\bl)=\chi(-\infty,\bl)=\chi'(\infty,\bl)\,.
\ee
This is again the matching property, analogous to \eqref{eq:matchingcond}.

We have seen that, as a consequence of the interplay between field equations and fall-off conditions, in the absence of external currents, the asymptotes of the scalar field at $\scri^+_+$ and $\scri^-_-$ and at $i^\pm$ vanish:
\begin{align*}
    \phi_+^{(1)}(+\infty) = \phi^{(1)}_-(-\infty)& =0\,,\\
    \lim_{\tau\to \infty}\phi\vert_{\Hcal_\tau^\pm} & = 0\,.
\end{align*}
Hence, with the area form on $S^2$ being proportional to $r^2$ in retarded coordinates, one shows that the corner term \eqref{cornerchargefromBVdual} is given by
\begin{equation}\label{asychfromBVdual}
    S^{\partial,\textrm{soft}}[\Gamma]\approx -\intl{\mathcal{I}^+_-} \lambda^{(1)}\phi_+^{(1)} d^2\Omega  + \intl{\mathcal{I}^-_+} \lambda^{(1)}\phi_-^{(1)} d^2\Omega \approx 0,
\end{equation}
which coincides with the conservation of the (smeared) asymptotic charge, as analysed in \cite{CC,CFHS}, with ${\lambda}^{(1)}$ an arbitrary function on the two dimensional celestial sphere.

In the notation of \cite{Her95}, we can write this as:
\[
-\int \lambda^{(1)}(\bl) \chi(-\infty,\bl)d^2\bl+ \int \lambda^{(1)}(\bl) \chi'(\infty,\bl) d^2\bl\approx 0\,,
\]
which is the smeared version of the matching property \eqref{MatchingProp:scalar} and we identify:
\[
Q^{{\rm soft}+}_{\lambda^{(1)}}\equiv -\intl{\mathcal{I}^+_-} \lambda^{(1)}\chi(-\infty,\bl) d^2\bl\,,\qquad
Q^{{\rm soft}-}_{\lambda^{(1)}}\equiv -\intl{\mathcal{I}^-_+} \lambda^{(1)} {\chi}'(+\infty,\bl) d^2\bl\,.
\]
Hence \eqref{asychfromBVdual} is the on-shell charge conservation:
\[
Q^{{\rm soft}+}_{\lambda^{(1)}}\approx Q^{{\rm soft}-}_{\lambda^{(1)}}\,.
\]

\subsection{Soft charge from the Fourier transform}
Another way to interpret formula \eqref{asychfromBVdual} uses the Fourier representation of the field, so can be applied only on Minkowski spacetime. In \cite{Her95} one writes the Fourier representation of the field $\ph(x)$ (denoted by $A(x)$ in the original) as
\[
\ph(x)=\frac{1}{\pi} \int a'(p) \delta(p^2)\epsilon(p^0) e^{-i x\cdot p} d^4p= \frac{1}{\pi} \int \frac{d^3\vec{p}}{2E_p} a'(E_p,\vec{p})e^{-i (x^0 E_p-\vec{x}\cdot \vec{p})}
\]
Let 
\[
a(\vec{p})\equiv a'(|\vec{p}|,\vec{p})\,,
\]
and we define (analogously to \cite{CCM}) the unsmeared soft charge as:
\be\label{df:charge}
Q^{\textrm{soft}+}_{\hat{x}}:=\lim_{\omega\rightarrow 0} \frac{\omega}{2}(a(\omega\hat{x})+a^\ddagger(\omega\hat{x}))\,.
\ee
Next, we note that
\[
a'(\omega \bl)=-\tilde{\dot\chi}(\omega,\bl)/\omega\,,
\]
and use thw following formula proven in \cite{Her95}:
\be\label{magic}
\widetilde{\dot\chi}(0,\bl)=\frac{1}{2\pi}\int_{-\infty}^{\infty}\dot\chi(s,\bl)ds=-\frac{1}{2\pi} \chi(-\infty,\bl)\,.
\ee
It is now easy to see that fields with non-vanishing $\chi(-\infty,\bl)$ are the \textit{infrared singular} ones ($1/\omega$ behavior around 0). Inserting this into \eqref{df:charge}, and identifying $\chi(-\infty,\bl)$ in retarded coordinates with $\phi_+^{(1)}(-\infty)$, we obtain:
\[
Q_{\hat{x}}^{{\rm soft}+}\sim \lim_{\omega\rightarrow 0}(\omega \tilde{\dot\chi}(\omega,\bl)/\omega+ c.c.)\sim \phi_+^{(1)}(-\infty,\hat{x})\,,
\]
so smearing with an arbitrary function $\lambda^{(1)}$ on the two-dimensional celestial sphere, we obtain
\[
Q_{\lambda^{(1)}}^{{\rm soft}+}\sim \int_{S^2} \lambda^{(1)}\phi
_+^{(1)} d^2\Omega\,,
\]
as expected.

\subsection{Hard charges for scalar fields}\label{Sec:hardscalar}
In this section we would like to approach the problem of computing hard contributions to the charge obtained in Section \ref{Sect:Asymduelmodel}, when sources for the scalar field are added to the model. We aim to utilise the dual model description to be able to gain information on asymptotic charges for scalar fields, however the duality outlined above strictly holds in the absence of sources. As a matter of fact, when a source for the scalar field is present, namely when\footnote{The superscript reminds us that $J^{(4)}_\phi$ is a top form, whereas $J_\phi$ is a function.} 
\be\label{sourcedscalarfieldeqt}
d\star d\phi = J_\phi^{(4)} = \star J_\phi,
\ee
we have $d\Hsf_\phi = J^{(4)}_\phi$, and the na\"ive duality outlined in Section \ref{Sect:Scal} breaks down (we defined $\Hsf_\phi\coloneqq\star d\phi$ as in Remark \ref{dualityremark}). 

\subsubsection{Duality in the presence of sources}
To extend the duality outlined above to the case of a scalar field coupled to external sources, we will consider a model encoding the equations of motion 
$$
    d\star d{\B} = \star J_{\B}.
$$
Let us specify this discussion for the case of Minkowski spacetime $\mathbb{M}$, for which $H^k(\mathbb{M}) = 0$, $k>0$. We wish to establish a duality between these two models, and we do so by parametrising the possible primitives of $d\star d\phi = J_\phi$ by means of the 2-form field $\B$, and the primitives of $d\star d\B = J_\B$ by means of the scalar $\phi$. Consider the following relations:
\begin{subequations}\begin{align}\label{dualmodels}
     d^\star d \phi &=J_\phi   &d^\star d\B &= J_{\B}\\\label{cohomologyparametrisation}
    \star d\phi &=d\B + \Hsf_\phi   &\star dB &= d\phi +\Hsf_\B
\end{align}\end{subequations}
with $\Hsf_\phi,\Hsf_\B$ choices of primitives\footnote{On manifolds with nontrivial cohomology, primitives are defined modulo closed forms, not necessarily exact. We will not discuss this case.}, i.e such that $d\Hsf_\phi=\star J_\phi$ and $d\Hsf_\B=\star J_\B$. By applying $d\star$ to the Equations in \eqref{cohomologyparametrisation}, we further derive the relations
\begin{equation}
    0 = d\star d\B + d\star \Hsf_\phi \qquad 0 = d\star d\phi + d\star \Hsf_\B.
\end{equation}
Then, if we want the models \eqref{dualmodels} to be ``dual'', we need to enforce
\begin{equation}\label{Sourcedduality}
    \begin{cases}
    -d^\star \Hsf_\phi = d^\star d\B \approx J_B\\
    -d^\star \Hsf_\B = d^\star d\phi \approx J_\phi
    \end{cases}
\end{equation}

\begin{remark}
Observe that the relations in \eqref{Sourcedduality} require that, if $\phi$ has no sources, $\Hsf_\phi$ must vanish, and $\B$ is ``dual'' only if also $J_\B=0$, since $d^*d\B=0$. This is an enhancement of the duality expressed by Equation \eqref{nonsourcedduality}, as $d^2\B=0$ implies that $\phi$ satisfies $\Box\phi=J_\phi$, and vice-versa. In particular, Equations \eqref{Sourcedduality} contain the standard sourceless duality \eqref{nonsourcedduality} as a special case.
\end{remark}

Then, if we define the interacting two-form model to be given by the action functional:
\be\label{Sourceddualmodel}
S_{\text{dual}}^{J}= \intl{M} d\B \star d\B  - \B \star J_{\B},
\ee 
with $d\star J_B=0$, we have that an interacting two-form model is related to an interacting scalar model whenever the relations \eqref{Sourcedduality} hold.

\begin{remark}
Observe that we do not need to know the explicit form of interaction that generates $J_{\B}$, and we can just consider it as an effective external source. The approach of \cite{CFHS} to hard scalar asymptotic charges, which proposes a link to a dual model with non-local sources, does not directly fit in our language. Although we were not able to simply adapt the argument used in their work, it would be interesting to understand how the two approaches might be related. We plan to address this question in our future work.
\end{remark}

\subsubsection{Calculation of hard charges} To compute the hard charge for this model, we modify the BV-BFV calculation of Section \ref{Sect:BVdualmodel} in the following way. The sourced classical action \eqref{Sourceddualmodel} is extended to a BV action functional in exactly the same way as we did in Section \ref{Sect:BVdualmodel}, with the difference that the BV operator $Q$ on the antifield $\B^\ddag$ will read 
\be
Q\B^\ddag = d\star d B - \star J_{\B} 
\ee
Hence, from Equation \eqref{dualboundaryaction} we get 
\begin{equation}
    S^\partial = \intl{M} d\gamma \left(d\star d\B - \star J_\B\right) + d\tau \B^\ddag
\end{equation}
Evaluating this on a gauge parameter $\Gamma\in\Omega^1(M)$, and in virtue of the (on-shell) relations \eqref{Sourcedduality} and \eqref{cohomologyparametrisation}, we have:
\begin{equation}
    S^\partial[\Gamma]\approx \intl{\partial M}d\Gamma \left( \star d\B + \star \Hsf_\phi\right) = \intl{\partial M} d\Gamma d\phi
\end{equation}
where now $\phi$ is a solution of $\Box \phi = J_\phi$.

Let us split $S^\partial[\Gamma]$ as in \eqref{EMFullBoundaryactionExplicit} and compute $S^{\partial,\textrm{soft}}_{\scri^+\cup\scri^-}[\Gamma]$ and $ S^{\partial, \textrm{hard}}_{\Hcal^+\cup\Hcal^-}[\Gamma]$ separately. We start with the soft charge:
\[
S^{\partial,\textrm{soft}}_{\scri^+\cup\scri^-}
    \approx 
    \intl{\scri^+\cup\scri^-}d\Gamma d\phi 
    = 
    -\intl{\mathcal{I}^+_-} \lambda^{(1)}\chi^{\textrm{out}}(-\infty,l) d^2\bl + \intl{\mathcal{I}^-_+} \lambda^{(1)} {\chi^{\textrm{in}}}'(+\infty,l) d^2\bl\,,
\]
since the free asymptotes at $\scri^{+}_-$ and $\scri^-_+$ are now $\chi^{\textrm{out}}(-\infty,l)$ and ${\chi^{\textrm{in}}}'(+\infty,l)$, respectively. They appear from the following decomposition of $\phi$, a solution to $\Box \phi=J_\phi$ (compare with \cite{Her95}):
\[
\phi=\phi^{\rm R}+\phi^{\rm in}=\phi^{\rm A}+\phi^{\rm out}.
\]
The free fields $\phi^{\rm in/out}$ solve the homogeneous equation \eqref{eq:free:scalar} and, assuming that incoming and outgoing fields satisfy the fall-off conditions \eqref{faloff1} and \eqref{faloff2}, we have the following identities for the asymptotes:
\begin{align*}
\chi(s,\bl)&=\chi_J(s,\bl)+\chi^{\rm in}(s,\bl)=\chi_J(+\infty,l)+\chi^{\rm out}(s,\bl)\\
\chi'(s,\bl)&=\chi_J(-\infty,l)+{\chi^{\rm in}}'(s,\bl)=\chi_J(s,\bl)+{\chi^{\rm out}}'(s,\bl)
\end{align*}
Hence
\[
\chi(+\infty,l)=\chi_J(+\infty,l)\,,\qquad \chi'(-\infty,l)=\chi_J(-\infty,l)\,,
\]
and we have the matching property (compare with \eqref{MatchingProp:scalar} in the free case):
\be\label{MatchingProp:scalar2}
\chi'(+\infty,l)=\chi(-\infty,l)\,,
\ee
We identify (in analogy to section \eqref{sec:Sym:ScalarField}):
\[
Q^{{\rm soft}+}_{\lambda^{(1)}}\equiv -\intl{\mathcal{I}^+_-} \lambda^{(1)}\chi^{\textrm{out}}(-\infty,l) d^2\bl\,,\qquad 
Q^{{\rm soft}-}_{\lambda^{(1)}}\equiv -\intl{\mathcal{I}^-_+} \lambda^{(1)} {\chi^{\textrm{in}}}'(+\infty,l) d^2\bl\,.
\]
The hard charge contribution is given by:
\begin{equation*}
    S^{\partial, \textrm{hard}}_{\Hcal^+\cup\Hcal^-}[\Gamma] \approx  \intl{\Hcal^+} d\Gamma d\phi -  \intl{\Hcal^-} d\Gamma d\phi
    = \intl{\Hcal^+\cup\Hcal^-} d\Gamma d (\Delta^{\textrm{A}}-\Delta^{\textrm{R}}) J_\phi = -\intl{S^2} d\Gamma  \chi_J \vert^{+\infty}_{-\infty}\,,
\end{equation*}
where (following \cite{Her95})
\[
\chi_J(s,\bl)=\int dy \delta(s-y\cdot \bl) J_\phi(y)
\]
and the boundary values are
\[
\chi_J(\pm\infty,l)=\intl{\Hcal^\pm} \frac{J_{\phi}(v)}{v\cdot \bl}  d\mu(v)\,,
\]
which agrees with \cite{CCM}, upon identification:
\[
Q^{{\rm hard}\pm}_{\lambda^{(1)}}\equiv \int -\lambda^{(1)}(\bl) \chi_J(\pm\infty,l)
\]
Hence
\[
S^{\partial, \textrm{hard}}_{\Hcal^+\cup\Hcal^-}[\Gamma] \approx Q^{{\rm hard}+}_{\lambda^{(1)}}-Q^{{\rm hard}-}_{\lambda^{(1)}}\,.
\]
Finally, we can write the formula for the boundary action in the form:
\begin{multline}\label{Total:charge:scalar}
S^\partial[\Gamma] \approx -\int \lambda^{(1)}(\chi^{\textrm{out}}(-\infty,l)+\chi_J(+\infty,l)) d^2\bl \\
+ \int \lambda^{(1)} ({\chi^{\textrm{in}}}'(+\infty,l)+\chi_J(-\infty,l)) d^2\bl\\=
-\int \lambda^{(1)}(\bl)\chi(-\infty,l)d^2\bl+\int \lambda^{(1)}(\bl)\chi'(+\infty,l)d^2\bl=Q^{+}_{\lambda^{(1)}}-Q^{+}_{\lambda^{(1)}}\approx 0\,,
\end{multline}
where 
\[
Q^{\pm}_{\lambda^{(1)}}=Q^{{\rm hard}\pm}_{\lambda^{(1)}}+Q^{{\rm soft}\pm}_{\lambda^{(1)}}\,.
\]
Hence \eqref{Total:charge:scalar} is the total charge conservation and at the same time, a smeared version of the matching property \eqref{MatchingProp:scalar2}, proven in \cite{Her95}.

\subsection{Shift symmetries by zero modes, global gauge transformations}\label{s:LGTnoBV}
In this section we would like to analyse a particular class of transformations that arise from considering either shifting a scalar field by a constant or, dually, the $\B$ field by an element {of $H^2(M)$ (assumed not empty)}. In \cite{CFHS} these are called \emph{large gauge transformations}, because in their work they are interpreted as ultimately being the same. We prefer to resort to the more standard nomenclature and refer to them as \emph{global gauge transformations}.

The action functional for a scalar field does not admit local gauge symmetries, but it admits shifts by constant maps
\begin{equation}
	\phi \longmapsto \phi + \alpha
\end{equation}
where $\alpha$ is a constant function on $M$, i.e. $d\alpha(x)=0$ or $\alpha\in H^0(M)$. 

Similarly, we have a symmetry for $\B$ generated by \emph{closed-but-not-exact} forms $\beta\in H^2(M)$, i.e. $d\beta=0$ but $\beta\not=d\gamma$. We observe, \emph{en passant}, that this is not possible on Minkowski space, since $H^2(\mathbb{M}^4)=0$.

Note that both these transformations are to be considered symmetries of the zero modes (more than a symmetry of the fields), i.e. elements of the kernel of the kinetic operator (that is the quadratic part of the Lagrangian density). We will see shortly that these transformations do not really admit a BV description in the usual sense.

Let us try to construct the BV-data for this field redefinition. The space of fields is now ($\alpha$ is promoted to ghost-number 1)
\begin{equation}
	\mathcal{F} = T^*[-1]\left( F_{cl} \times H^0[1](M)\right)
\end{equation}
and the extended BV action reads
\begin{equation}
	S^{BV}_{\text{large}}=\intl{M} \star d\phi \wedge d\phi + \phi^\ddag\alpha,
\end{equation}
with $\phi^\ddag$ the \emph{cotangent} field for $\phi$.

The $-1$-shifted symplectic BV-form is 
$$\Omega=\int \delta \phi\delta\phi^\ddag + \delta\alpha\delta \alpha^\ddag\,,$$
and the action of the BV operator $Q$ on fields is
\[
	Q\phi = \alpha;\quad   Q\phi^\ddag = d\star d\phi;\quad Q\alpha=0;\quad Q\alpha^\ddag = \phi^\ddag
\]
so that $\iota_Q\Omega = \delta S + \pi^*\alpha^\partial$. On the other hand, the BV extension for large symmetries in the case of the dual model reads
$$
    S^{BV}_{\text{dual,large}}=
        \intl{M} d\B\star d\B + \B^\ddag \beta
$$
and the associated BV operator
\begin{eqnarray*}
    Q\B = \beta & Q\beta=0 \\
    Q\B^\ddag = d\star d\B & Q\beta^\ddag = \B^\ddag
\end{eqnarray*}

The problem with the above na\"ive construction is that these operators are not coboundaries, i.e. $Q^2\not=0$. In fact, we compute
$$
    Q^2\alpha^\ddag = Q (Q\alpha^\ddag) = Q \phi^\ddag = d\star d \phi\not=0,
$$
and similarly for the dual model:
$$
    Q^2\beta^\ddag = d\star d\B\not=0.
$$
both of which only vanish on shell. Hence this construction (for symmetries given by constants and, dually, elements of the second cohomology group) \emph{does not yield a BV data}. 

Ignoring this and pushing through with formal calculations, for the shift $\B \to \B + \beta$ with $\beta\in H^2(M)$, one gets a ``formal boundary action''
\be\label{rigidscalartransformations}
S^\partial_{\Wcal_R}[\beta] \approx \intl{\partial M} \beta d\phi = \intl{\partial M} d(\beta \phi)
\ee
which is a corner term. In the limit $R\to \infty$, assuming the same fall-off $\beta = \sum_{k=1}^\infty r^{-k} \beta^{(k)}$ we get that $d\beta=0$ implies that $\beta^{(k)} = d\beta^{(k+1)}$. In particular $\beta^{(0)} = 0 =d \beta^{(1)}$ and $\beta^{(1)}= d\beta^{(2)}$. Then, equation \eqref{rigidscalartransformations} becomes
\begin{align*}
    S^\partial_{\scri}[\beta]= & \lim_{R\to \infty} S^\partial_{\Wcal_R}[\beta] = \intl{\scri} d(\beta^{(1)} \phi^{(1)}) \\
    = & \intl{\scri} d(d\beta^{(2)} \phi^{(1)}) = \int_{S^2} d\beta^{(2)} \phi^{(1)}\vert^{+\infty}_{-\infty} = \int_{S^2} \mathrm{dVol}_{S^2}\widetilde{\lambda} \phi^{(1)}\vert^{+\infty}_{-\infty}.
\end{align*}
This is the same conclusion as the one reached in \cite{CFHS}. It is evident, though, that on spaces with trivial second cohomology, the procedure used in \cite{CFHS} needs to be better understood. One possibility, might be to phrase this in terms of relative cohomology (see, e.g., \cite{BottTu}). We note, however, that the construction presented in Sections \ref{Sect:BVdualmodel} and \ref{Sect:Asymduelmodel} avoids this problem, while still reproducing the correct asymptotic behaviour. Thus, we believe it provides a neat description of how soft charges emerge from the symmetries of the dual model.

One possible way to overcome the difficulty above might be to think of the global transformations for the scalar/dual model as shifts in zero modes, rather than proper symmetries. One can extend the BV construction in order to consider infinitesimal shifts in the space of zero modes, a framework that is related to \emph{formal geometry}. In a nutshell, introducing a differential $\mathbb{d}$ on the space of solutions to $d\star d B = 0$, which we can think of being a field-version of de Rham differential, we obtain that, for $\beta\in \Omega^2(M)_{\text{coclosed}}$
\be
    S_{\text{formal}} = \intl{M} d(B + \beta) \star d(B + \beta) + B^\ddag \mathbb{d}\beta
\ee
satisfies the \emph{differential modified classical master equation}:
\be
    \{S_{\text{formal}},S_{\text{formal}}\} = d S_{\text{formal}}^\partial
\ee
reconstructing a formal version of the BV-BFV construction. We refer to \cite[Section 3]{BCM} and \cite[Section 3.3.2]{CMW} for an introduction of this technique in relation to the Poisson sigma model and, more generally, AKSZ theories, and defer its analysis for the case at hand to a subsequent work. \footnote{We would like to thank K. Wernli for pointing this out to us.}

\section*{Acknowledgements}
This research was (partly) supported by the NCCR SwissMAP, funded by the Swiss National Science Foundation. MS acknowledges partial support from Swiss National Science Foundation grants \verb|P2ZHP2_164999| and \verb|P300P2_177862|. KR acknowledges the support of EPSRC through the grant \verb|EP/P021204/1|. Both authors would like to thank the Perimeter Institute, where part of this research was completed, for hospitality and the opportunity to work in such an inspiring research environment. We would also like to thank A.~Ashtekar, M.~Campiglia, A.S.~Cattaneo, L.~Freidel and A.~Herdegen for enlightening discussions.

\bibliographystyle{alpha}
\bibliography{AsymptoticSymmetriesBV-BFV}

\end{document}